\documentclass[12pt,preprint,rotating]{aastex}
\newcommand{\xmm}{{\it XMM-Newton} }
\newcommand{\asca}{{\it ASCA} }
\newcommand{\sax}{{\it BeppoSAX} }
\newcommand{\chandra}{{\it Chandra} }

\slugcomment{submitted to ApJ}

\shorttitle{Ni abundance in the core of the Perseus Cluster}
\shortauthors{Gastaldello \& Molendi}

\begin{document}
   
\title{Ni abundance in the core of the Perseus Cluster: an answer to the significance of resonant scattering}

%% Use \author, \affil, and the \and command to format
%% author and affiliation information.
%% Note that \email has replaced the old \authoremail command
%% from AASTeX v4.0. You can use \email to mark an email address
%% anywhere in the paper, not just in the front matter.
%% As in the title, you can use \\ to force line breaks.

\author{Fabio Gastaldello\altaffilmark{1}}
\affil{IASF - CNR, via Bassini 15, I-20133 Milano, Italy}

\email{gasta@mi.iasf.cnr.it}

\and

\author{Silvano Molendi}
\affil{IASF - CNR, via Bassini 15, I-20133 Milano, Italy}
\email{silvano@mi.iasf.cnr.it}

\altaffiltext{1}{Universit\`a di Milano Bicocca, Dip. di Fisica, P.za della Scienza 3 I-20133 Milano, Italy} 

\begin{abstract}

Using an \xmm observation of the Perseus cluster  
we show that the excess
in the flux of the 7-8 keV line complex previously detected by \asca and \sax 
is due to an overabundance of Nickel rather than to an anomalously high 
Fe He$\beta$/Fe He$\alpha$ ratio. 
This observational fact leads to the main result that resonant 
scattering, which was assumed to be responsible for the supposed anomalous 
Fe He$\beta$/Fe He$\alpha$ ratio, is no longer required. The absence of 
resonant scattering points towards the presence of significant gas motions
(either turbulent or laminar) in the core of the Perseus cluster. 

\end{abstract}

%% Keywords should appear after the \end{abstract} command. The uncommented
%% example has been keyed in ApJ style. See the instructions to authors
%% for the journal to which you are submitting your paper to determine
%% what keyword punctuation is appropriate.


%%%%%%%%%%%%%%%%%%%%%%
\section{Introduction}
%%%%%%%%%%%%%%%%%%%%%%

The X-ray emission from clusters is due to a diffuse, tenuous (with typical 
densities of $10^{-4}-10^{-2}\,\rm{cm}^-3$), and hot (with typical 
temperatures of $10^{7}-10^{8}\,\rm{K}$) thermal plasma. Although for 
these ranges of density and temperature the gas is optically thin
to Thomson scattering for the continuum, it can be optically thick in the
resonance X-ray lines of highly ionized atoms of heavy elements 
\citep{gilfanov87}.
Apart from other interesting observable effects \citep{sazonov02}, the major
effect of resonance scattering (the absorption of a line photon followed by 
immediate re-emission) is to distort the surface-brightness profile
of the cluster in the resonance line due to diffusion of photons from the
dense core into the outer regions of the cluster. This must be taken into
account when attempting to determine element abundances from X-ray 
spectroscopic observations of galaxy clusters. In fact only with the key
assumption that the plasma is optically thin lines equivalent widths can 
unambiguously convert to element abundances, when fitting CCD spectra with 
the available plasma codes. In presence of resonant 
scattering the true abundances in the core of clusters are significantly
underestimated because the line emission is attenuated due to 
photons scattered out of the line of sight. To make things worse, 
the most promising
line for resonant scattering is the He$\alpha$ Fe emission line at 6.7 keV
\citep{gilfanov87,sazonov02}
which is also one of the most prominent emission line in cluster spectra 
and in general drives the global abundance determination. 

High sensitivity and high resolution spectrometers are needed 
to directly measure the spectral features of resonant scattering (as 
modification 
of the line profile or resolution of the He-like line into its constituents in 
order to determine directly the effects of scattering) and polarimeters, 
to detect the polarized scattered radiation \citep{costa01,sazonov02}.
Currently 
the simplest method to reveal and estimate the presence of resonance scattering
is to compare the fluxes of an expected 
optically thick line and of an optically thin 
one and to check if it is correctly modeled by a plasma code assuming optically
thin emission. This was done in the past with \asca and \sax 
for the ratio between
He$\alpha$ Fe line at 6.7 keV and the He$\beta$ Fe line at 7.90 keV 
(which is expected to have an optical depth typically 
smaller than one for resonant scattering) and in
particular the best data were the ones for the Perseus cluster.

\citet{mole98} analyzed data collected with the MECS on board \sax and
found that the ratio of the flux of the 7-8 keV line complex to the 6.7 keV
line was significantly larger than predicted by optically thin plasma code
and that the ratio decreases with increasing cluster radius. They 
noted that this effect could be explained either by resonant scattering
or by a Ni overabundance, eventually favoring the former explanation.
On the contrary \citet{dupke01}, according to the 
experimental evidence of a central enhancement of SNIa ejecta in cD clusters, 
favored the over-abundant Ni explanation. 

These were the two hypothesis that the resolution and sensitivity of past 
instruments could not resolve. \xmm has now for the first time 
the combination of resolution and effective area at high energies to give 
an unambiguous answer to the question. In this paper our aim is to try to 
solve this controversy. A complete spectral and spatial analysis, 
in particular of the temperature structure of the Perseus cluster which 
requires a detailed temperature map, is beyond the scope of this paper. 

The outline of the paper is as follows. In section 2 we give information about
the \xmm observation and data preparation. In section 3 we 
present spatially resolved measurements of temperature and
Ni and Fe abundances. In section 4 we discuss our results and draw our 
conclusions.

At the nominal redshift of Perseus (z=0.0183), 1 arcmin corresponds to
22.2 kpc ($H_0 =$ 70 km s$^{-1}$ Mpc$^{-1}$,
$\Omega_{\rm m} = 1 - \Omega_{\Lambda} =$ 0.3).
In the following analysis, all the quoted errors are
at $1 \sigma$ (68.3 per cent level of confidence)
unless stated otherwise.

%%%%%%%%%%%%%%%%%%%%%%%%%%%%%%%%%%%%%%%%%%%
\section {Observation and Data Preparation}
%%%%%%%%%%%%%%%%%%%%%%%%%%%%%%%%%%%%%%%%%%%

The Perseus cluster was observed with \xmm \citep{jansen01} during Revolution
210, with the THIN1 filter and in Full Frame Mode, for 53.6 ks for MOS and 51.2
 ks for PN, but resulting in an effective exposure time (as written in the 
keyword LIVETIME of the fits event file) of 53.1 ks for the MOS and 24.7 ks for
 the PN. We generated calibrated event files using the publicly available 
 SASv5.3.3.

To fully exploit the excellent EPIC data from extended and low 
surface brightness objects and from this observation in particular, 
the EPIC background needs to be correctly taken into account.
The EPIC background can be divided into a cosmic X-ray background, dominant 
below 2-3 keV, and an instrumental background, dominant for energies higher
than 2-3 keV (for what concern the continuum emission, apart from fluorescence 
lines). This latter component can be further divided into a detector
noise component, present in the low energy range (below 300 eV) and in a 
particle induced background, which is the major concern for our scientific 
case.
The particle induced background consists of a flaring component, characterized
by strong and rapid variability, produced by soft protons (with energies 
less than few hundreds of keV) which are funneled towards the detectors by the 
mirrors, and a second more stable component associated with high energy 
particles interacting with the structure surrounding the detectors and the
detectors themselves. The latter component has been studied using CLOSED
filter observation, shows only small intensity variations and is characterized
by a flat spectrum and a number of fluorescent lines. Apart from a rather
strong variability of the fluorescent lines this component can be 
properly subtracted using a large collection of background data.
The common way to face the flaring component is to remove periods of high
background, because the S/N is highly degraded, especially at high energy 
(where the data are crucial to measure the exponential cut-off and thus the
temperature of the emitting plasma) and because the shape of the 
spectrum is varing with 
time \citep{arnaud01}. The strategies to reject these flaring periods are
mainly two: selection of time intervals where the count rate in a given 
high energy band is lower than a given threshold 
(which has been our approach in previous analysis where we fixed the thresholds
at 0.35 cts/s for PN in the 10-13 keV band and 0.15 cts/s in the 10-12 keV band
for MOS, based on Lockman Hole data) or finding  a mean count 
$\overline{\rm{c}}$ and
then choosing as a threshold value $\overline{\rm{c}}+3\sigma$, by means 
of a Gaussian or
Poissonian fitting or $\sigma$-clipping methods (see Appendix A 
of Pratt \& Arnaud 2002 
for the second approach and Marty et al. 2002 for a general discussion on
soft proton cleaning criteria).

\begin{figure}
\epsscale{0.8}
\rotate
\plotone{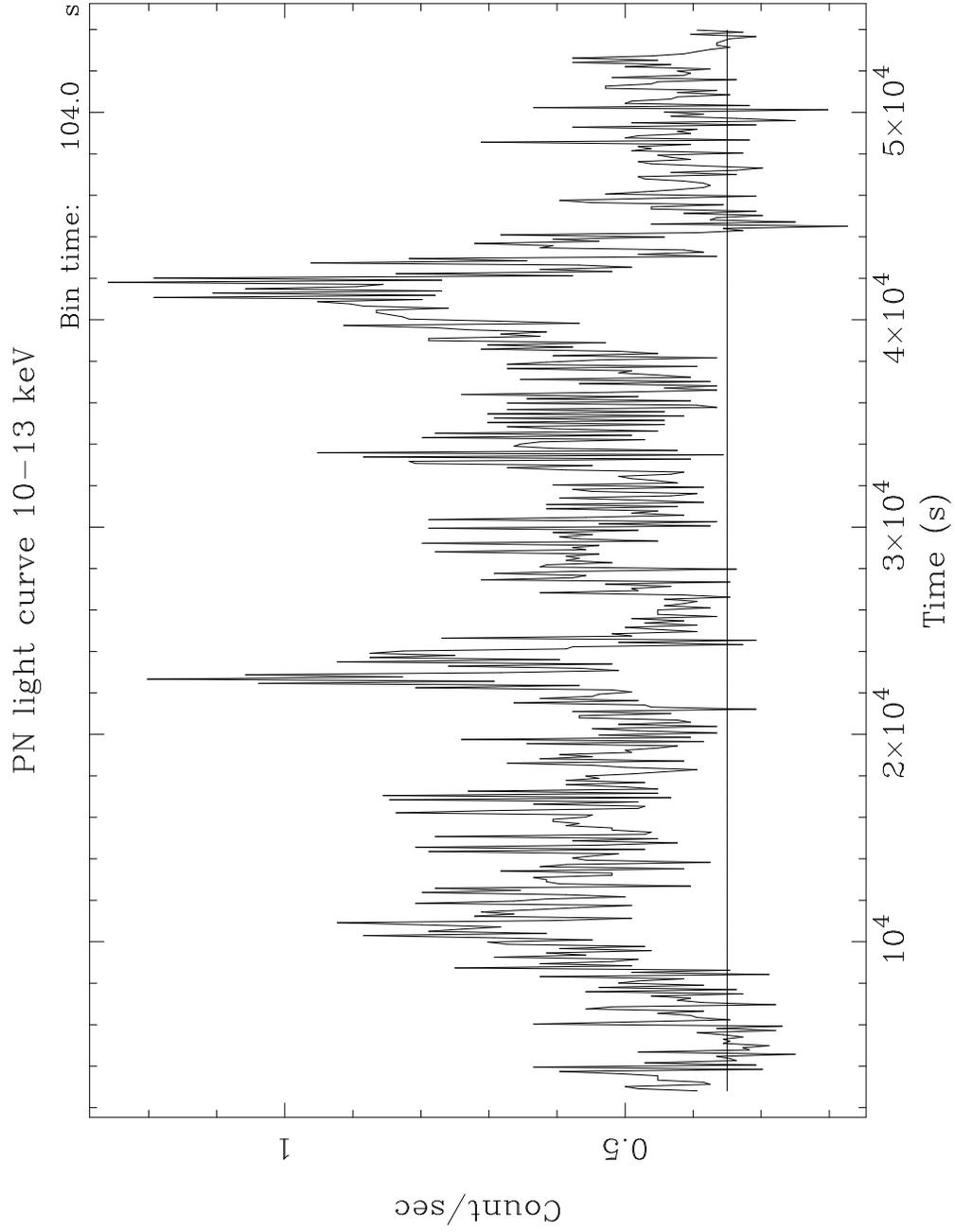}
\caption
{PN light curve in the energy band 10-13 keV together with our threshold of
0.35 cts/s. 
\label{lcurve}}
\end{figure}

The light curve in the 10-13 keV band for the PN observation of the Perseus 
cluster is shown in Fig.\ref{lcurve} together with our standard threshold of
0.35 cts/s. It is evident that the observation is badly affected by 
soft proton and if we adopt our threshold all the observation would 
be rejected.
The light curve is also structured in such a way 
that a 3$\sigma$-clipping method 
rejects only 529 s of observation, finding a mean rate of 0.53 cts/s with a 
standard deviation of 0.15 cts/s,
while fitting with a Gaussian and rejecting all the intervals above $3\sigma$ 
from the mean rejects only 700 s of observation.   
Our approach was therefore to consider all the observation for two reasons:
we can exploit the fact that Perseus is the brightest X-ray cluster and it is
so bright in its central zone that the background, also in presence of a 
high level of soft protons as we have in our observation, is not important;
moreover we can try to model the soft proton which contaminate 
the spectra using
in first approximation a power law as a background model (which means that the
model is not convolved via the effective area of the instrument). The
self-consistency and viability of our approach will be shown in the results.
     
We have accumulated spectra in 9 concentric annular regions centered on the
emission peak with bounding radii $0.5^{\prime}-1^{\prime}$, 
$1^{\prime}-2^{\prime}$, $2^{\prime}-3^{\prime}$, $3^{\prime}-4^{\prime}$, 
$4^{\prime}-5^{\prime}$, $5^{\prime}-6^{\prime}$, $6^{\prime}-8^{\prime}$, 
$8^{\prime}-10^{\prime}$, $10^{\prime}-14^{\prime}$. We did not consider
the inner bin inside $0.5^{\prime}$ in order to avoid contamination by the 
power law spectrum of the Seyfert cD galaxy NGC 1275. 
\newline
Spectra have been accumulated for the three cameras independently and the 
blank fields provided by the calibration teams were used as background 
\citep{lumb02}.  
Background
spectra have been accumulated from the same sky regions as the source 
spectra, after
reprojection onto the sky attitude of the source (this ensures the proper 
subtraction in the same way as it was performed in detector co-ordinates, see 
Lumb 2002). 
\newline
The vignetting correction has been applied to the effective area generating 
effective area files for the different annular regions using the 
SAS task \emph{arfgen}. We generate flux weighted arf using exposure corrected
 images of the source as detector maps and the parameter \emph{extended source}
switched to true, following the prescription of \citet{saxton02}. Spectral 
results for the cluster A3528 obtained
in this way and with the vignetting correction applied directly to the spectra
 \citep{arnaud01}
are practically the same \citep{gasta03}. 
We also correct the PN spectra for out of time events following 
the prescriptions of \citet{grupe01}.
The redistribution
matrices used are m1\_r6\_all\_15.rmf (MOS1),  m2\_r6\_all\_15.rmf (MOS2) and,
 depending
on the mean ``RAWY'' of the region, the set of ten single-pixel matrices, from 
epn\_ff20\_sY0.rmf to epn\_ff20\_sY9.rmf, and double-pixel matrices, from 
epn\_ff20\_dY0.rmf to epn\_ff20\_dY9.rmf, for PN.
\newline
Due to its higher effective area (further increased by the use of 
doubles data) and similar spectral 
resolution at high energies, the PN camera 
will be the leading instrument in our analysis and 
the one for which the results are most compelling, 
in particular for what concerns the Ni abundance. 
There are still some problems for what concern the three 
EPIC cameras cross-calibration and in particular
at high energies the study of power-law sources returns
harder spectra for MOS1, intermediate for MOS2 and then the softest for PN 
\citep{kirsch02}. 
Also our analysis of the galaxy
cluster A3528 gives systematically higher temperatures and abundances
 for MOS1 respect to MOS2 and PN.
The conclusions of a recent work aimed at assessing the EPIC spectral 
calibration using a simultaneous \xmm and \sax observation of 3C273 
strengthen this fact:
the MOS-PN cross calibration has been achieved to the available statistical 
level except for the MOS1 in the 3-10 keV band which returns flatter spectral 
slope \citep{mole03}.  

%%%%%%%%%%%%%%%%%%%%%%%%%%%%%%%%%%%%%%%%%%%%%%%%%%
\section{Spectral modeling and energy ranges used}
%%%%%%%%%%%%%%%%%%%%%%%%%%%%%%%%%%%%%%%%%%%%%%%%%%

All spectral fitting has been performed using version 11.2.0 of the XSPEC 
package \citep{arnaud96}.  

As a first step we concentrate on the hard band which is the one of
interest to determine the abundances of iron and nickel and also to make a
direct comparison with the MECS results. 
We use three different energy 
bands: 3-10 keV, 3-7 keV in order to have a band less contaminated by 
the hard tail of 
soft protons, and the 3-13.5 keV and 3-12 keV for PN and MOS respectively 
in order to have more data to acceptably model the soft protons background. 
\newline
When fitting the first two bands we analyze the spectra 
with a single temperature VMEKAL model \citep{mewe85,kaastra92,liedahl95} 
with the 
multiplicative component WABS to account for the Galactic absorption
 fixed at the value
of $0.143 \times 10^{22}\,\rm{cm^{-2}}$ (according to Schmidt et al. 2002). 
We leave  the abundances of Ar, Ca, Fe and Ni
(the only elements which have emission lines in the range 3-10 keV) free 
and keep all the other abundances fixed to half the solar values 
\citep{fukazawa00} (this corresponds for example to the 1T (3-10 keV) model in Tab.~\ref{pntab} and Tab.~\ref{m2tab}).
\newline  
When fitting the wider high energy band, more contaminated by soft protons,
 we add a power law background model 
(VMEKAL+POW/B in XSPEC) in order to model the soft proton background component 
(this corresponds to the 1T+pow/b (3-13.5 keV) model in Tab.~\ref{pntab} and 
1T+pow/b (3-12 keV) Tab.~\ref{m2tab}).

As a second step we fit the entire energy band 0.5-10 keV with two models:
\newline
a single temperature model leaving $\rm{N_{H}}$ to vary freely 
(the fit is substantially improved respect to the one with $\rm{N_{H}}$ fixed 
to the galactic value) and the abundance of 
O, Ne, Mg, Si, S, Ar, Ca, Fe and Ni. For the outer annuli, when 
required from the previous analysis in the hard band, we add the pow/b 
component with normalization and slope fixed at the best fit values found (these models corresponds to 1T (0.5-10 keV) or 1T+pow/b (0.5-10 keV) in Tab.~\ref{pntab} and Tab.~\ref{m2tab});
\newline
a two temperature model ( WABS*(VMEKAL+VMEKAL) in XSPEC) where the metal 
abundance of each element of the second thermal component is bound to be 
equal to the same parameter of the first thermal component. As for the single
temperature model we add the pow/b component when required (these models corresponds to 2T (0.5-10 keV) or 2T+pow/b (0.5-10 keV) in Tab.~\ref{pntab} and Tab.~\ref{m2tab}). 
\newline 
The two temperature model is a rough attempt to reproduce the complex spectrum
resulting from projection effects, azimuthal mean of very different emission
regions (like holes and luminous regions in the Perseus cluster, see 
Schmidt et al. 2002; Fabian et al. 2002) and an atmosphere probably 
containing components at different temperatures, as in M87 
\citep{kaiser03,mole02}. 

We also allow the redshift to be a free parameter in order to 
account for any residual gain calibration problem.
We adopt for the solar abundances the values of \citet{grevesse98},
where Fe/H is $3.16\,\times\,10^{-5}$. 
To make comparison with previous measurements, a simple rescaling can 
be made to obtain the values with the set of abundances of \citet{anders89},
where the solar Fe abundance relative to H is $4.68\,\times\,10^{-5}$ 
by number.

%%%%%%%%%%%%%%%%%
\section{Results}
%%%%%%%%%%%%%%%%%

%%%%%%%%%%%%%%%%%%%%%%%%%%%%%%%%%%%%%%%%%%%%%%%
\subsection{1T results in the high energy band}
%%%%%%%%%%%%%%%%%%%%%%%%%%%%%%%%%%%%%%%%%%%%%%%

\begin{figure}
\epsscale{1.0}
\plotone{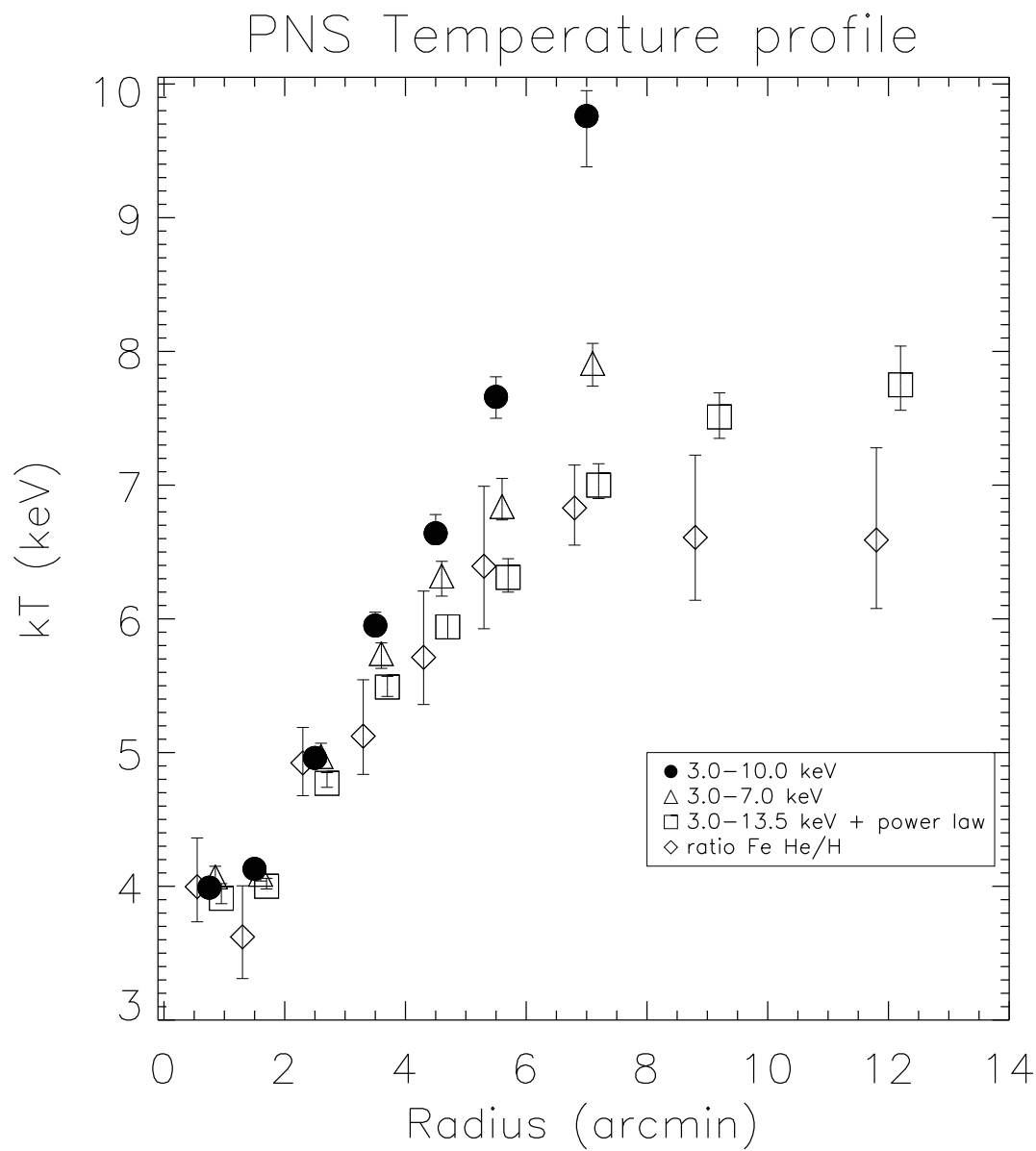}
\caption
{PN singles temperature profile. Uncertainties are at the 68\% level for one 
interesting parameter($\Delta \chi^{2}\,=\,1$). Full circles represent the 
temperature obtained using the range 3-10 keV, open triangles the temperature 
obtained using the range 3-7 keV and open squares
the temperature obtained by using the range 3-13.5 keV and adding to the 
source model a power law
background component. In the last two bins we do not show the temperatures 
obtained in the 3-10 keV and 3-7 keV bands because they are larger than 
10 keV. 
Diamonds represent the temperature obtained by the ratio of the fluxes of 
He$\alpha$ to H$\alpha$ Fe lines. 
\label{tpns}}
\end{figure}

In Fig.\ref{tpns} we show the temperature profile obtained 
analyzing the single events spectrum for the
PN camera. This is also an example of our working procedure.  
The full circles refer to the results obtained
using the 3-10 keV band, while the open triangles indicates the 
results obtained using the 3-7 keV band
with the Ni abundance frozen to the best fit value obtained in 
the 3-10 keV band. It is clear that 
where the source is overwhelmingly bright the hard 
component of the soft proton does not affect 
the spectrum and there are no differences between the temperatures 
obtained in different energy bands, 
while in the outskirts of the cluster, where the source brightness 
is lower and the
soft protons become important, the fitted plasma temperature 
reaches uncorrect and unphysically
high values and large residuals at high energy are present.
With the open squares we show the temperature obtained by fitting not only 
the source but 
also the soft protons with a power law background model, in the energy 
band 3-13.5 keV: as we expect in the inner region adding
the background component does not affect the temperature determination 
nor statistically improve the fit,
on the contrary in the outer regions the temperature are 
significantly reduced and the fit is
improved, eliminating the residuals at high energies. For example in 
the $6^{\prime}-8^{\prime}$ ring
the simple single temperature fit gives a $\chi^{2}$ of 1502 for 
1137 degrees of freedom, while the fit
with the power law background model in addition gives a 
$\chi^{2}$ of 1258 for 1288 d.o.f.

To confirm our results we compare the 
temperatures obtained in this way with those
obtained with the MECS instrument on board \sax \citep{grandi02a}. 
The temperature profiles, apart from
the differences in the three camera due to the cross-calibration 
problems we discuss before (confirmed also with the superb statistics 
of Perseus), are in good agreement at least up to 8 arcmin. In the outer rings 
between 8 and 14 arcmin the increasing
importance of background relative to source counts prevent 
us from recovering a correct temperature 
with our method (see De Grandi \& Molendi 2002a for a more general discussion 
about \xmm and \sax temperature 
determinations and the greater sensitivity of the latter over the former 
to low surface brightness regions due to much lower background). 

With a determination of the temperature structure we can 
address the issue of metal abundances measure
and attempt to discriminate between the 
presence of resonant scattering or the supersolar abundance of Nickel. 
Resonant scattering is 
increasingly important towards the center of the cluster so 
we choose our two inner bins to test its presence.
Fitting the spectra with a MEKAL model, assuming solar ratios, 
actually does not reproduce the 
8 keV line complex. As shown in the left panel of 
Fig.\ref{nores} for the $1^{\prime}-2^{\prime}$ bin, the emission is 
underestimated as for previous missions (see Fig.1 of Molendi et al. 1998 
for example).  
However the data show for the first time that the excess 
is due
to an uncorrect modeling of the Ni He$\alpha$ line complex at 7.75-7.80 keV 
(in the rest frame of the source) and not to an underestimation of the Fe 
He$\beta$ line which is correctly modeled. Infact if we fit the data 
with a VMEKAL model, we eliminate almost completely the residuals 
and give a better fit with a Ni 
abundance of 1.23 in solar units, as shown in the right panel of 
Fig.\ref{nores}. The fit with a MEKAL model gives a $\chi^{2}$ of 855 for 802
d.o.f for the $0.5^{\prime}-1^{\prime}$ bin and 1116 for 1023 d.o.f. for the 
$1^{\prime}-2^{\prime}$ bin, while 
a fit with a VMEKAL model (with Ar and Ca fixed to 0.5 $Z/Z_{\odot}$, because 
they are not important in driving the fit, in order to have only the Ni 
abundance as additional free parameter) gives a $\chi^{2}$ of 835 for 
801 d.o.f. for the first bin and 1092 for 1022 d.o.f. for 
the second bin, with $\Delta\chi^{2}$ which are statistically significant 
at more
that the 99.9\% according to the F-test (the value of the F statistics is 
F=19.2 with a probability of exceeding F of $1.4\times10^{-5}$ for the first 
bin and F=22.5 with a probability of exceeding F of $2\times10^{-6}$ for the 
second bin).

We can conclude that the ratio of He$\beta$/He$\alpha$ Fe lines is not 
anomalously high respect to the optically thin model and that it is 
not necessary to invoke resonant scattering in the core of the 
Perseus cluster. The excess in the flux in the 8 keV line complex 
respect to a MEKAL model is entirely due to Ni overabundance with respect 
to solar values, as was previously suggested \citep{dupke01}. 

The reader will notice some residuals in the He$\alpha$ Fe line 
complex at 6.7 keV. This is an instrumental artifact present only in the inner
bins out to $2^{\prime}$ of the PN camera we suspect connected to some 
residual CTI problems due to the 
high flux of the Perseus cluster. The net effect is to lower the energy 
resolution broadening the line profile. We test that this does not affect our 
results fitting spectra for our two inner bins with a bremsstrahlung 
model plus two Gaussians fixed at
the energies of the Fe He$\alpha$ at 6.67 keV and Fe He$\beta$ at 7.90 keV 
leaving the redshift, width and normalizations of the two lines as free 
parameters. We find that the Gaussian width of the He$\alpha$ line in the 
two bins are $5.3\times10^{-2}$ keV in the $0.5^{\prime}-1^{\prime}$ bin and 
$4.5\times10^{-2}$ keV in the $1^{\prime}-2^{\prime}$ bin. If we force the 
He$\beta$ to have up to a width of $8\times10^{-2}$ the excess due to the Ni 
He$\alpha$ line blend at 7.75-7.80 keV is still significantly present.
This instrumental effect is evident because of the large equivalent width of
 the Fe line at 6.67 keV and does not alter significantly the measure of metal
 abundances as we show further on.

We can make some other important 
considerations investigating another line ratio, namely 
the He$\alpha$ Fe line complex at 6.7 keV
 over the H$\alpha$ Fe line at 6.97 keV. This ratio allows a robust and 
independent determination of the temperature, because as the temperature 
increases the contribution from the
He Fe line decreases while the contribution from the H Fe line increases. 
Thus the intensity ratio of
the two lines can be used to estimate the temperature. This was done in 
the past determining the 
variation with the temperature of the centroid of the blend of the two lines, 
because gas proportional counters did not have sufficient spectral resolution 
to resolve the two lines \citep{mole99}. Now with \xmm
we can resolve the lines, measure separately their intensity and use 
their ratio as a thermometer.
To do that we obtain a calibration curve of the line flux ratio as a 
function of temperature simulating spectra with MEKAL model and the PN 
singles response matrix with a step size of 0.1 keV, fixing 
the metal
abundance of 0.3 solar units and the normalization to unity in XSPEC units 
(however the flux ratio is independent from these two quantities), with an 
exposure time of 100 ks to ensure
negligible statistical errors. We then model the spectra with a 
bremsstrahlung model plus two Gaussians 
for the two iron lines, in the energy range 3-10 keV and obtaining the 
fluxes of the two lines from
the best fit models. We obtain a calibration curve identical to that of 
\citet{neva03}.
We then measure the line flux ratio from the cluster PN singles data 
using the energy range 5.0-7.2 keV
to minimize the dependence from the continuum and calibration 
accuracy and to better describe the lines. 
We fitted each
spectrum with a bremsstrahlung model plus two Gaussians 
(using ZBREMSS plus two ZGAUSS models in XSPEC)
leaving all the parameters free, included the redshift (to take into account 
any possible gain calibration problem), except the 
line energies. 
The fits for all the annular bins
were good with a reduced $\chi^{2}$ never worse than 1.1 and the 
results for the temperature derived 
from line flux ratio are plotted as diamonds in Fig.\ref{tpns}. 
\newline  
As we can see also this independent temperature determination 
is in good agreement with all the others at
least out to 3 arcmin where the cluster is very bright and in 
good agreement out to 8 arcmin with
the measurement obtained from the model with the power law 
background component, confirming the validity of our modeling.
In the last two bins the temperature derived from
the lines ratio agrees well with the MECS measurement and starts to differ 
from the determination with power law background component, 
pointing to the fact that our modeling is not sufficient to fully take 
into account the background in these bins where the source is too dim
compared to the soft proton background. We can conclude that our temperature 
determination is reliable out to 8 arcmin.
\newline
The concordance between lines ratio and continuum temperature 
determination adds another piece of 
evidence against resonant scattering. In fact
since the Fe H$\alpha$ line optical depth is 1.8 times smaller than the Fe 
He$\alpha$ one (this is the difference in their oscillator strength), if 
resonant scattering is present, we would expect the ratio of 
He$\alpha$/H$\alpha$ lines to be lower than in the optically thin case. 
In turn this would lead to an 
overestimate of the temperature. Since this is not the case we can 
conclude that resonant scattering is not present.  

In Fig.\ref{feni} we plot the abundance profiles of Fe and Ni
 determined by our best fit model (thermal model plus power law component 
for the soft proton background, in the 3-13.5 keV band for PN and 3-12 keV for
MOS). 
We find an evident gradient in both elements: for Fe it agrees well 
with previous determinations, as the \sax one (without considering the 
corrections for resonant scattering, as done in Molendi et al. 1998), 
while we have for the first time a detailed abundance gradient
for Ni, with measurements reliable out to 8 arcmin (we show only the PN data 
as we discussed before). We stop at this radius because at larger radii
the temperature determination is no longer reliable and strong emission lines
of Ni, Cu and Zn induced by particle events affect the spectrum in the 
crucial range 7.5-8.5 keV \citep{freyberg02}.  
It is evident that there are some problems with the iron determination by MOS1,
as we also found in A3528.
 
Knowing that the excess in the 8 keV line complex is due to the Ni line, we 
can go back to BeppoSAX-MECS data and fit them with a VMEKAL
model allowing Ni abundance to be free. We find an abundance profile in 
agreement with the more detailed \xmm one, as shown in Fig.\ref{feni}.      

%%%%%%%%%%%%%%%%%%%%%%%%%%%%%%%%%%%%%%%%%%
\subsubsection{Results with the APEC code}
%%%%%%%%%%%%%%%%%%%%%%%%%%%%%%%%%%%%%%%%%%

We try to cross-check the results obtained with the MEKAL code with the
ones obtained with the APEC code \citep{smith01}. \citet{churazov03} 
pointed out that the APEC code has different energies for the Ni He$\alpha$ 
line complex, resulting in a different fitting for the line.
For comparison with Fig.\ref{nores} we show the fit of the 
$1^{\prime}-2^{\prime}$ bin with the recently released APEC version 1.3.0 
in Fig.\ref{apec} (but we have to notice that differences in the redshift determination 
could play a role too: the fit with MEKAL found a redshift of $1.49\times10^{-2}$, while 
the fit with APEC found a redshift of $1.54\times10^{-2}$). 
The excess is still present, although 
the statistical improvement of a variable Ni abundance respect to ratios 
fixed at solar values is not as evident as in the MEKAL case: the fit with an
APEC model gives a $\chi^{2}$ of 844 for 802
d.o.f for the $0.5^{\prime}-1^{\prime}$ bin and 1138 for 1023 d.o.f. for the 
$1^{\prime}-2^{\prime}$ bin, while 
a fit with a VAPEC model gives a $\chi^{2}$ of 843 for 
801 d.o.f. for the first bin and 1124 for 1022 d.o.f. for 
the second bin (with a probability of exceeding the F statistic of $3.8\times10^{-4}$ for the second bin).
These results are not conclusive, because there are additional issues with 
APEC, even in this latest release: forbidden and inter-combination lines of the
He-like Ni triplet are missing and, even after adding these lines, the total
Ni emission is still underestimated, thus worsening the differences between
APEC and MEKAL and making the excess less evident 
(see \verb!http://cxc.harvard.edu/atomdb/issues_caveats.html!).

%%%%%%%%%%%%%%%%%%%%%%%%%%%%%%%%%%%%%%%%%%%%%%%%%%%%%
\subsection{1T and 2T results in the 0.5-10 keV band}
%%%%%%%%%%%%%%%%%%%%%%%%%%%%%%%%%%%%%%%%%%%%%%%%%%%%%

We fit one and two temperature models to MOS2 and PN single data (we avoid
MOS1 data for the calibration problems explained in the previous section)
in the full energy band 0.5-10 keV. 
Single temperature models cannot adeguatly fit the entire band spectra,
giving temperatures systematically lower than the ones obtained in the hard 
band and leaving large residuals at high energies. These facts hint towards
the presence of more than one temperature component, in-fact 
a two temperature model yields a substantially better fit than the one
temperature model, although it is still not statistically acceptable, as the
reduced $\chi^{2}$ shows in Tab.~\ref{pntab} and Tab.~\ref{m2tab}. 

The temperature profiles for PN singles data and MOS2 are shown in Fig.~\ref{t_2t}: the two temperature fit shows the presence of a hot and of a cold 
component.
The temperature of the hot component especially in the outer bins matches 
the temperature determined with the fit in the hard band, while the temperature
of the cold component is less constrained, being about 2 keV in the PN fit 
and oscillating between 2 and 3 keV in the MOS2 fit. The relative
normalization of the two components, shown in Fig.\ref{em}, shows that the 
cool component is stronger in the center of the cluster, as we expect for cool
core clusters. Although there are some puzzling results, as in the inner two
bins of the MOS2 data where the fitting procedure prefers to give more 
importance to the cool component, and the presence of cool emission also in 
the outer bins where the emission should be negligible (although the cooling 
radius for Perseus is $\sim 6$ arcmin, Peres et al. 1998).

The Fe abundance profile, shown in Fig~\ref{fe2t} is not substantially changed
and in particular the abundance gradient is even more evident. Instead for the
Ni abundance profile, shown in Fig~\ref{ni2t}, the evidence for a gradient is 
not present. In fact adding a linear component improves the 
fit ($\chi^{2}=1.9$ for 5 d.o.f) respect to a constant 
($\chi^{2}=1.9$ for 6 d.o.f) for the PN Ni abundances derived by the 1T model
in the hard band, while the Ni profile derived by the 2T model in the entire 
band is essentially flat (fitting a constant returns a $\chi^{2}=3$ for 6 d.o.f 
and a linear component does not improve the fit, $\chi^{2}=2.9$ for 6 d.o.f).  
We caution the reader that the 2T 
modelization is rather complex, because some not completely justified 
assumptions are made, as for example that the abundances of the two components 
are equal, and there is some degeneracy in the contribution of the two 
X-ray emission components and the soft proton power-law background (see  
for MOS2 and PN the substantial difference in the temperature of the 
cool component \footnote{With the latest release of SAS, version 5.4.1, which 
revise the quantum efficiency for the MOS and the PN, the agreement between the 
two detectors should be better, in particular in the low energy band.}). Therefore the derived Ni abundance 
should be taken with some 
caution. Moreover it is very difficult to explain, in presence of a confirmed
Fe abundance gradient, a flat Ni abundance profile and a Ni/Fe ratio which 
increases going outward.

%%%%%%%%%%%%%%%%%%%%
\section{Discussion}
%%%%%%%%%%%%%%%%%%%%

Our main result can be summarized as follows: there is no need to invoke 
resonant scattering in the 
Fe He$\alpha$ line in the Perseus cluster core and the Fe abundance 
determination with optically thin emission models is reliable.

Resonant scattering should be important in the core of galaxy clusters, this 
is particularly true for
the Fe He$\alpha$ line in the core of the Perseus cluster 
(it has an optical depth of 3.3 
according to Sazonov et al. 2002). The optical depth of a resonance line 
depends on the characteristic 
velocities of small scale internal motions, which could seriously diminish 
the depth $\tau$ (Gilfanov et al. 1987, see Mathews et al. 2001 for an 
example of a detailed 
calculation in presence of turbulent motions). The absence of a 
clear evidence of resonant scattering strongly points towards the presence 
of significant gas motions. 
In fact, following \citet{gilfanov87}, the optical depth is

\begin{equation}
\tau = \tau^{0}\,(1 + v_{turb}^{2}/v_{Fe}^{2})^{-1/2}
\end{equation}

where $\tau^{0}$ is the optical depth at the line center in the absence of 
turbulence (in spectroscopy the word 
turbulence is used for all hydrodynamic motions of unknown pattern which 
cause a broadening of the spectral lines. In hydrodynamics turbulence has a 
much more restricted meaning), $v_{turb}$ is the turbulent velocity of the gas and $v_{Fe} = (2kT_{e}/M_{Fe})^{1/2} \sim c_{s}/8.8$ is the 
thermal speed of the iron ions and $c_{s}$ denotes the adiabatic sound speed 
in the ICM. Thus the absence of resonant scattering, $\tau < 1$, and assuming 
$\tau^{0} = 3.3$ \citep{sazonov02}, implies gas motions with 
characteristic velocities greater than $0.36\,c_{s}$, i.e a Mach number 
$M \gtrsim 0.36$.

Studies of optical line emission in the 
central regions of ellipticals reveal chaotic gas kinematics typically 
about 0.2-0.4 of the sound speed in the hot gas \citep{caon00} and
since small, optically visible line-emitting regions at 
T $\sim 10^{4}$ K are likely to be strongly  
coupled to the ambient gas, as some models predict \citep{sparks89} and 
clear correlation 
between H$\alpha$+[N II] and X-ray luminosities suggests, the hot gas 
should share the same turbulent velocities.  
The first clear example of resonant scattering, acting on the 
$2p-3d$ line of Fe XVII at 15.0 \AA\/ (0.83 keV),  
has been recently found in the giant elliptical galaxy 
NGC 4636 \citep{xu02} using the reflection grating on board \xmm and 
measuring the cross dispersion profile of the ratio between an optically thin 
emission blend, the two $2p-3s$ lines of Fe XVII at 17.0-17.1 \AA\/ 
(0.73 keV) and the optically thick line at 15.0 \AA\/. 
\citet{xu02} found that if an average turbulent velocity 
dispersion more than 1/10 of the sound speed is added to the assumed Maxwellian
the model becomes incompatible with the ratio of the 17.1 \AA\//15.0 \AA\/ 
lines. 
We remind that the detection of the resonant scattering is
only in the inner $1^{\prime}$, in-fact the phenomenon does not affect spectra 
extracted within a full-width of $2^{\prime}$ \citep{xu02}. Another elliptical,
NGC 5044, was 
observed with the RGS \citep{tamura03} and no evidence of resonant scattering
was found in spectra extracted in the full $2^{\prime}$. 
If also in this case a cross dispersion analysis where to show resonant 
scattering, these would rule out possible associations at least 
at these inner scales with optically line-emitting gas, because NGC 4636 
and NGC 5044 are the most striking examples of chaotic gas 
kinematics in the sample of \citet{caon00}.

Another possible source of gas motions is the activity of an AGN, which is 
now thought to be widespread in the core of galaxy clusters and 
strongly related to hot bubbles, for which one of the best cases is indeed 
the Perseus cluster \citep{fabian02}. The induced motions could be either 
turbulent or laminar, as suggested by the recent \chandra and optical results 
in \citet{fabian03a,fabian03b} (see the discussion about the flow causing 
the horseshoe $\rm{H}\alpha$ filament and the derived velocity of 700 $\rm{km\,s^{-1}}$ which for a sound speed of about 1170 $\rm{km\,s^{-1}}$, for a temperature of 5 keV, implies $M \sim 0.6$ or about the sound waves generated by the 
continuous blowing of bubbles).  
AGN activity could explain the 
lack of resonant scattering also in the other best candidate M87 
(but see also the discussion suggesting caution 
for these interpretation in the analysis of RGS data for M87 of 
Sakelliou et al. 2002).
\newline
What is becoming progressively clearer is that resonant scattering 
effects must be small and confined on small inner scales. 

The Fe abundance gradient confirms the general picture of an 
increase of SNIa ejecta in the center of relaxed cD clusters. The 
Ni abundance, because Ni is almost exclusively produced by SNIa, 
and the presence of 
a gradient also in this element could be a crucial confirmation of this
general picture (see De Grandi \& Molendi 2002b which report measures 
of Fe and Ni for 
a sample of 22 clusters observed with \sax and in particular their Fig.6 
showing a segregation between relaxed cD clusters and not relaxed clusters, 
with the formers with greater Fe and Ni abundances respect to latters).
The Ni abundance gradient is evident in the fit in the high energy band 
and, looking back at the \sax data, we can attribute the excess in the 8 keV 
line complex to an increased Ni abundance. However the complex thermal 
structure of the gas prevents us from reaching a robust determination 
of the Ni abundance profile. Detailed temperature and abundances maps are required to address this issue.

\acknowledgments
S. De Grandi, A. De Luca, F. Pizzolato, E. Churazov and F. Brighenti 
are thanked for useful discussion and the referee for important suggestions. 
S. De Grandi is also thanked for kindly providing MECS Ni measurements.
This work is based on observations obtained with \emph{XMM-Newton}, an ESA 
science mission with instruments and contributions directly funded by 
ESA Member States and the USA (NASA).

\clearpage

%--------------------------------- references

%\clearpage

%% Use the figure environment and \plotone or \plottwo to include 
%% figures and captions in your electronic submission.

\clearpage

\clearpage

\begin{figure}
\plotone{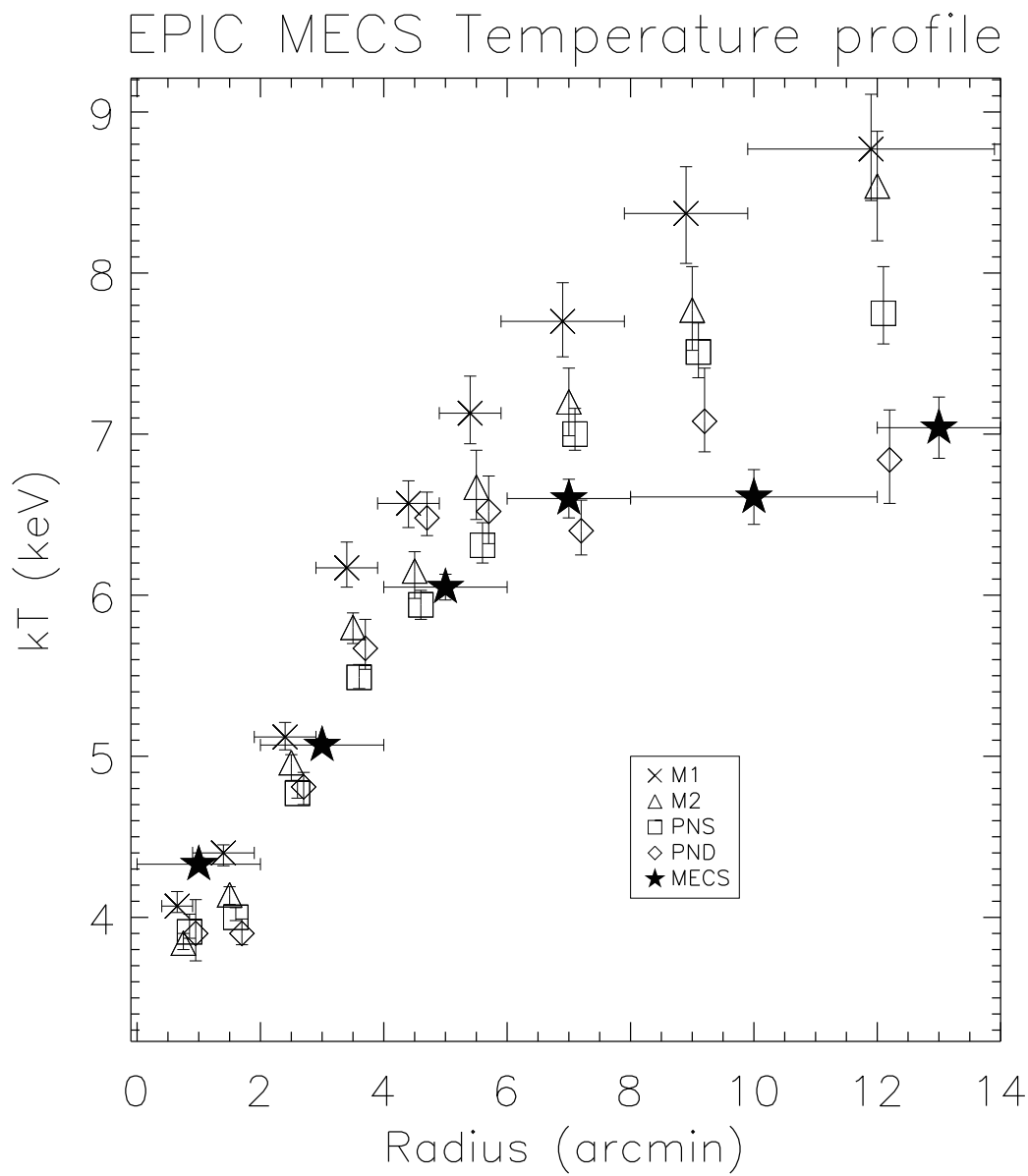}
\caption
{Temperature profiles obtained with the various EPIC cameras compared with the 
temperature profile obtained with the MECS instrument on board \sax. 
Uncertainties are at the 68\% level for one interesting parameter
($\Delta \chi^{2}\,=\,1$). Crosses represent temperatures obtained with MOS1, 
open triangles with MOS2, open squares with PN singles, open diamonds with
PN doubles and full stars with the MECS.
\label{tepic}}
\end{figure}

\clearpage

\begin{figure}
\plottwo{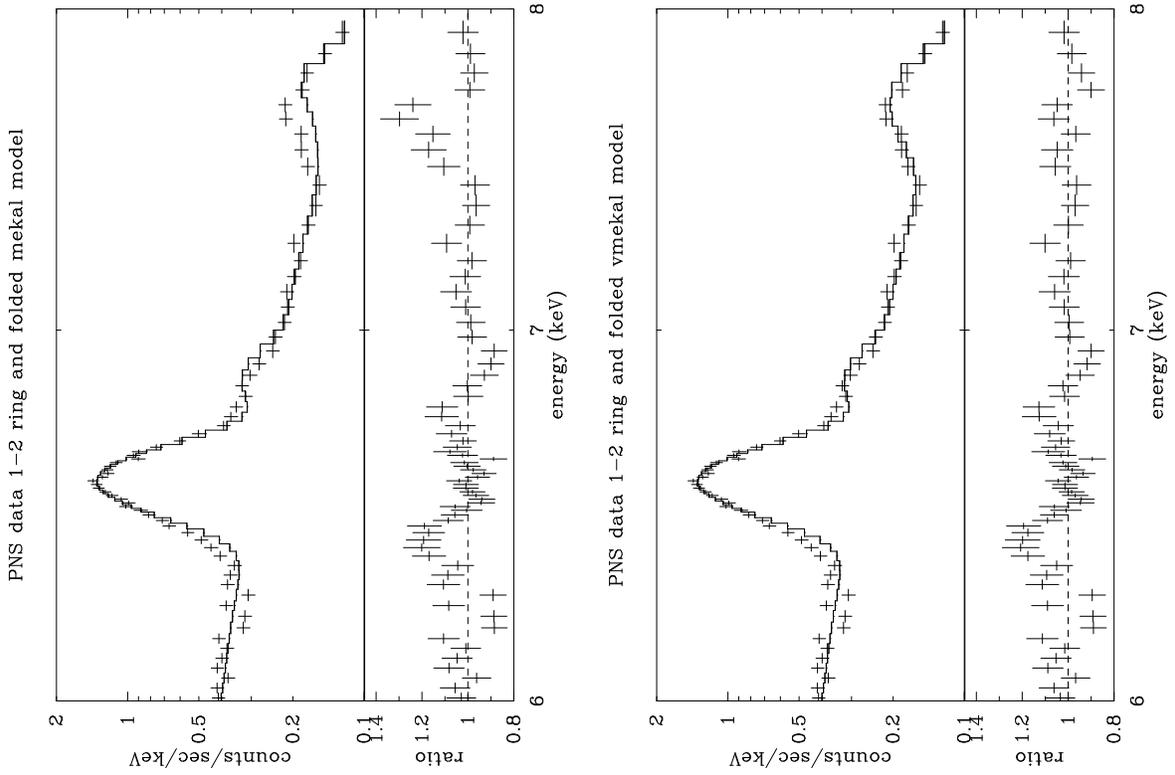}{f4b.eps}
\caption
{PN singles data for the $1^{\prime}-2^{\prime}$ bin in the 6-8 keV band and 
the corresponding fit with a MEKAL model, 
on the left, and with a VMEKAL model, on the right, together with the 
corresponding ratios of data respect to models.
\label{nores}}
\end{figure}

\clearpage

\begin{figure}
\epsscale{1.3}
\plottwo{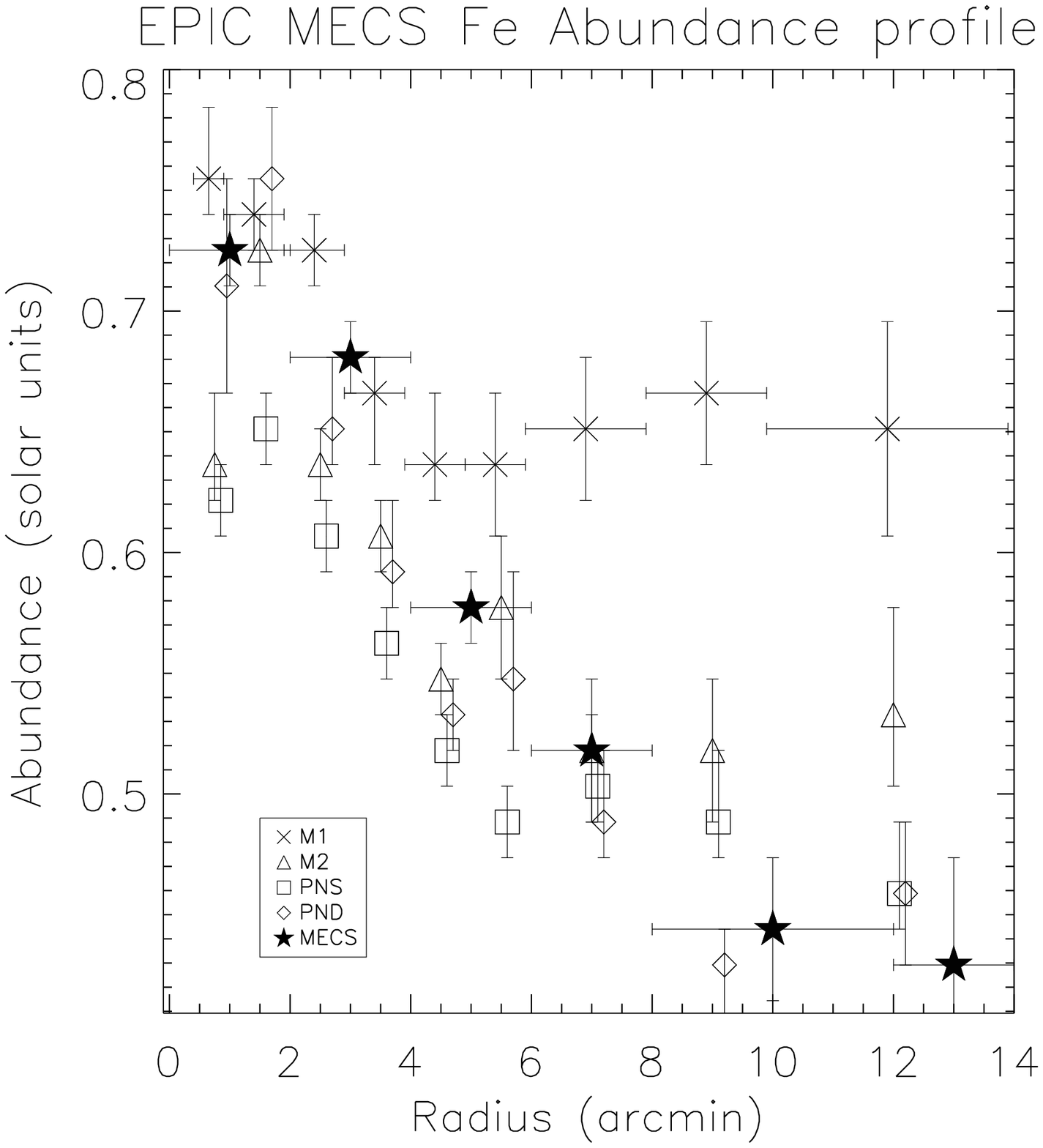}{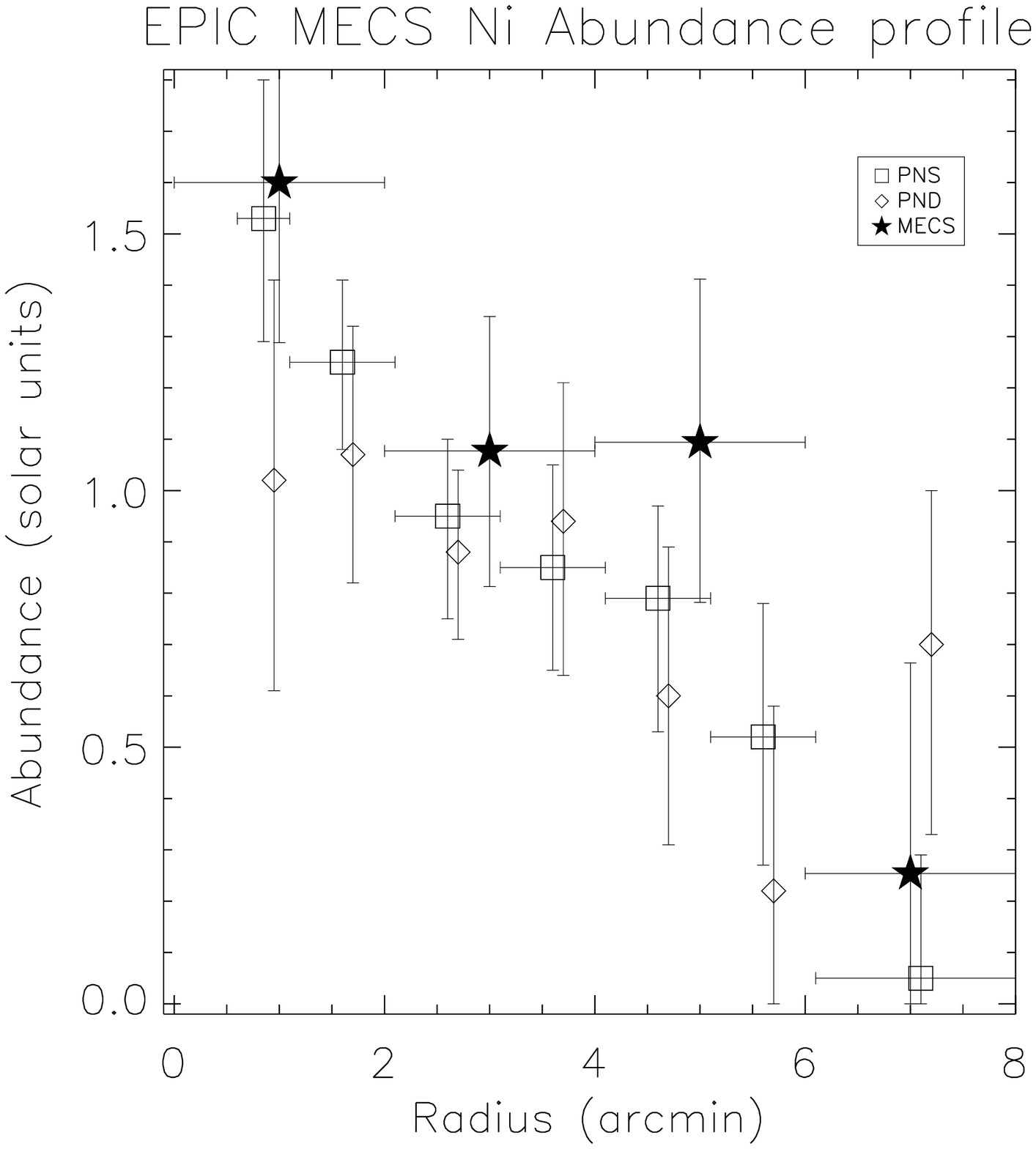}
\caption
{Abundance profiles for Fe, on the top, and for Ni, on the bottom, obtained 
with the various EPIC cameras compared with the 
abundance profiles obtained with the MECS instrument on board \sax. Uncertainties are at the 68\% level for one interesting 
parameter ($\Delta \chi^{2}\,=\,1$).
As in Fig.\ref{tepic}, crosses refers to MOS1, open triangles to MOS2, open
squares to PN singles, open diamonds to PN doubles and full stars to MECS. 
\label{feni}}
\end{figure}

\clearpage

\begin{figure}
\epsscale{0.5}
\plotone{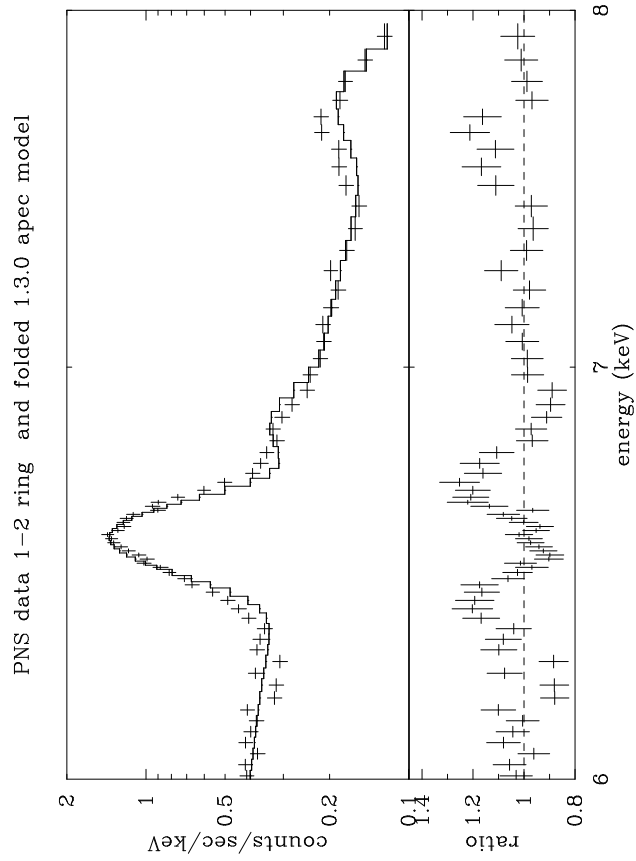}
\caption
{PN singles data for the $1^{\prime}-2^{\prime}$ bin in the 6-8 keV band and 
the corresponding fit with an APEC model version 1.3.0, 
together with the corresponding ratios of data respect to models.
\label{apec}}
\end{figure}

\clearpage

\begin{figure}
\epsscale{1.3}
\plottwo{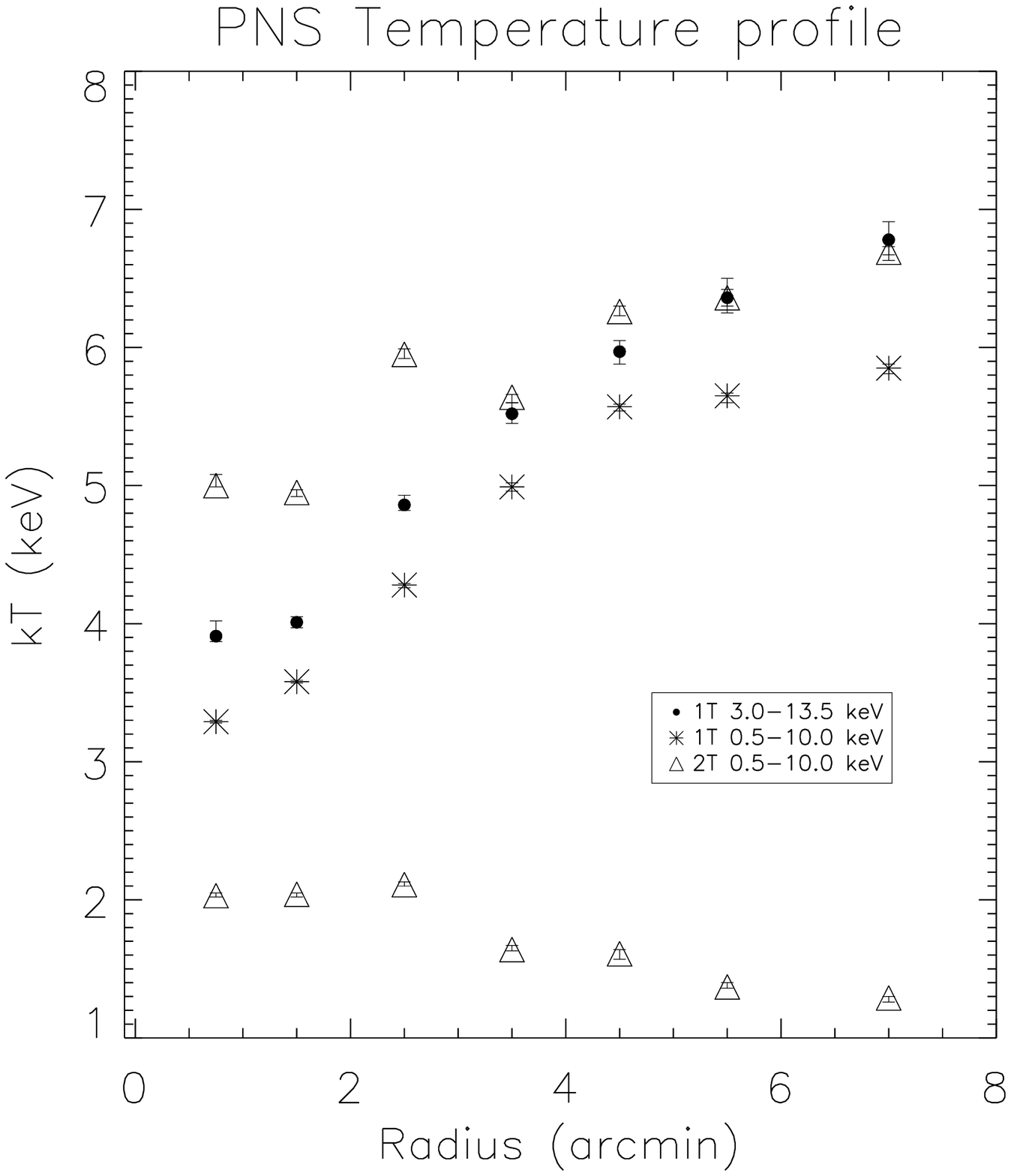}{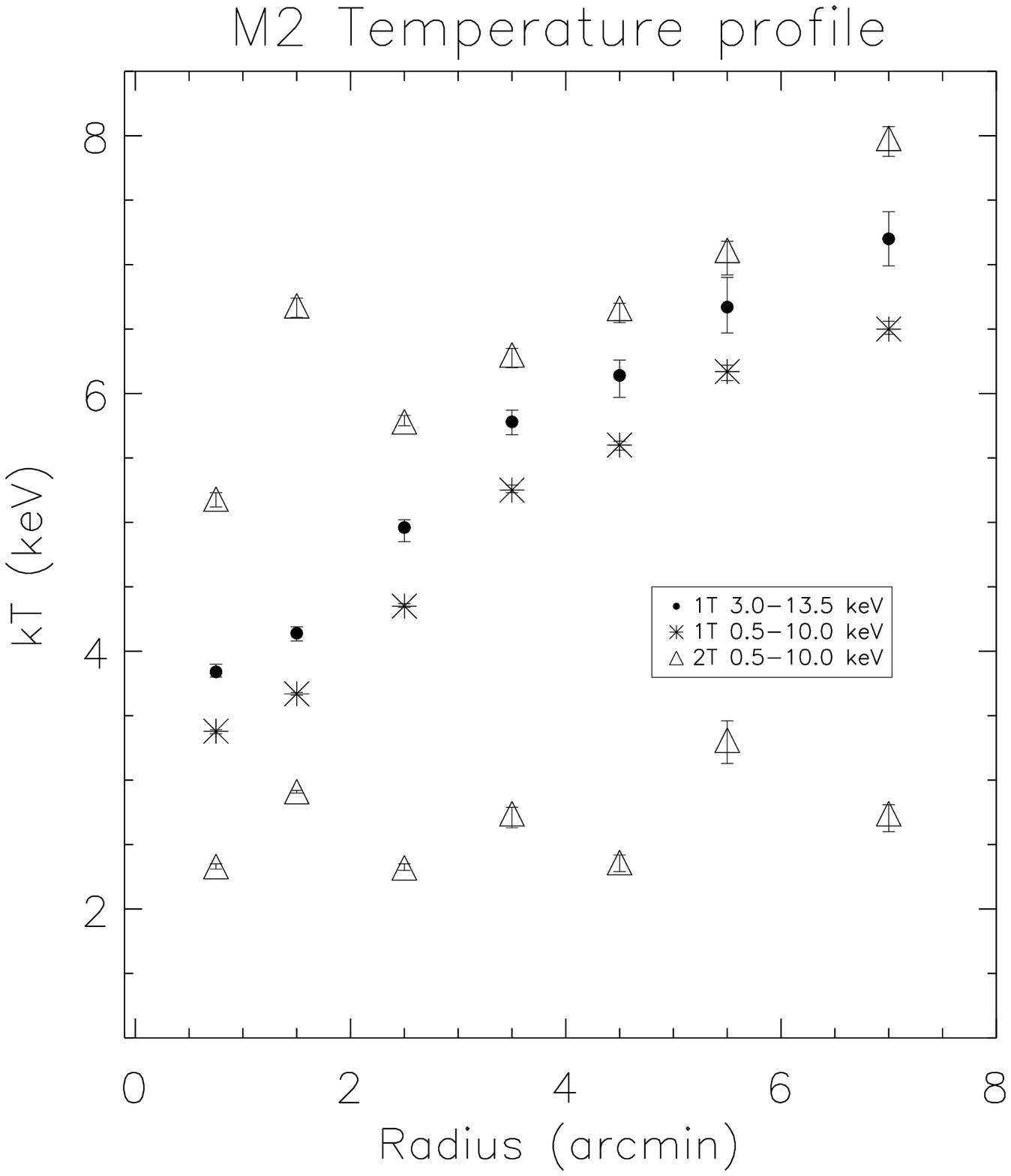}
\caption
{Temperature profiles obtained in the entire 0.5-10 keV band using one temperature (crosses) and two temperature models (open triangles, both the temperatures of the hot and cold component are shown) and, for comparison, the one obtained with the best fit model in the hard band (full circles), for PN singles data, on the top, and for MOS2 data, on the bottom. Uncertainties are at the 68\% level for one interesting parameter ($\Delta \chi^{2}\,=\,1$).\label{t_2t}}
\end{figure}

\clearpage

\begin{figure}
\epsscale{1.0}
\plotone{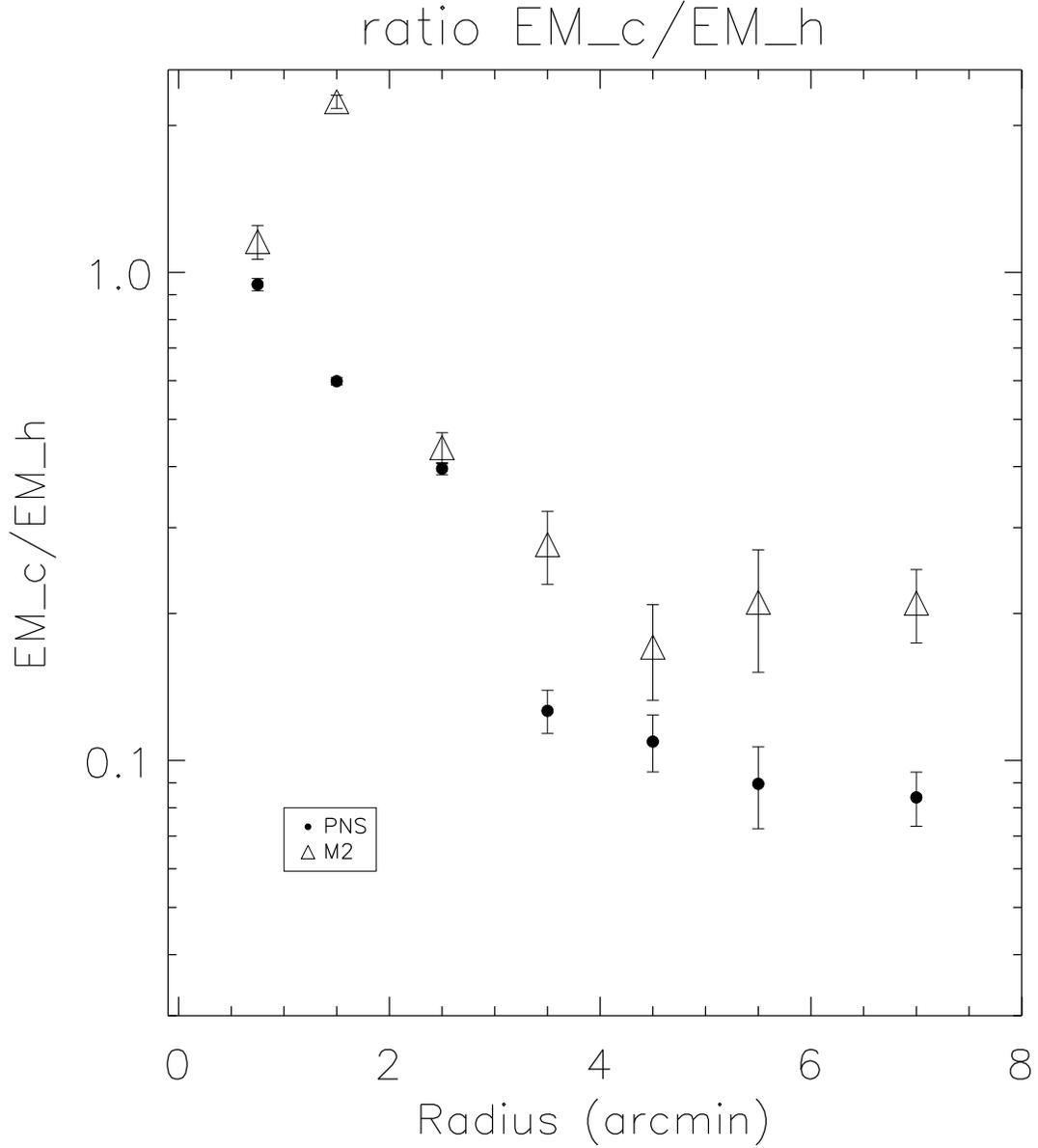}
\caption
{Ratio of the normalizations of the two temperature components, obtained by a two temperature model fitted in the 0.5-10 keV band, for PN single data (full circles) and MOS2 data (open triangles). Uncertainties are at the 68\% level for one interesting parameter ($\Delta \chi^{2}\,=\,1$). \label{em}}
\end{figure}

\clearpage

\begin{figure}
\epsscale{1.3}
\plottwo{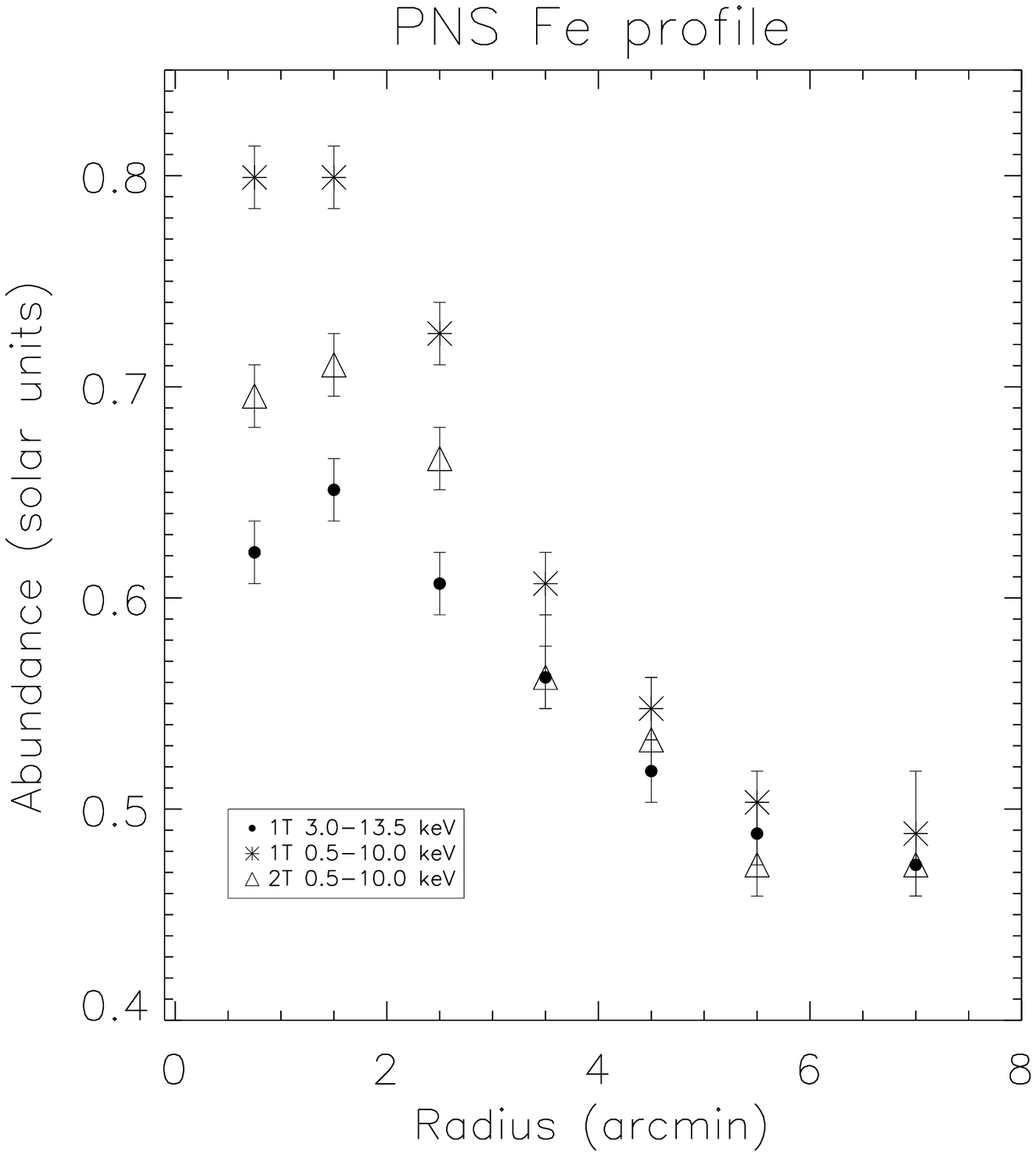}{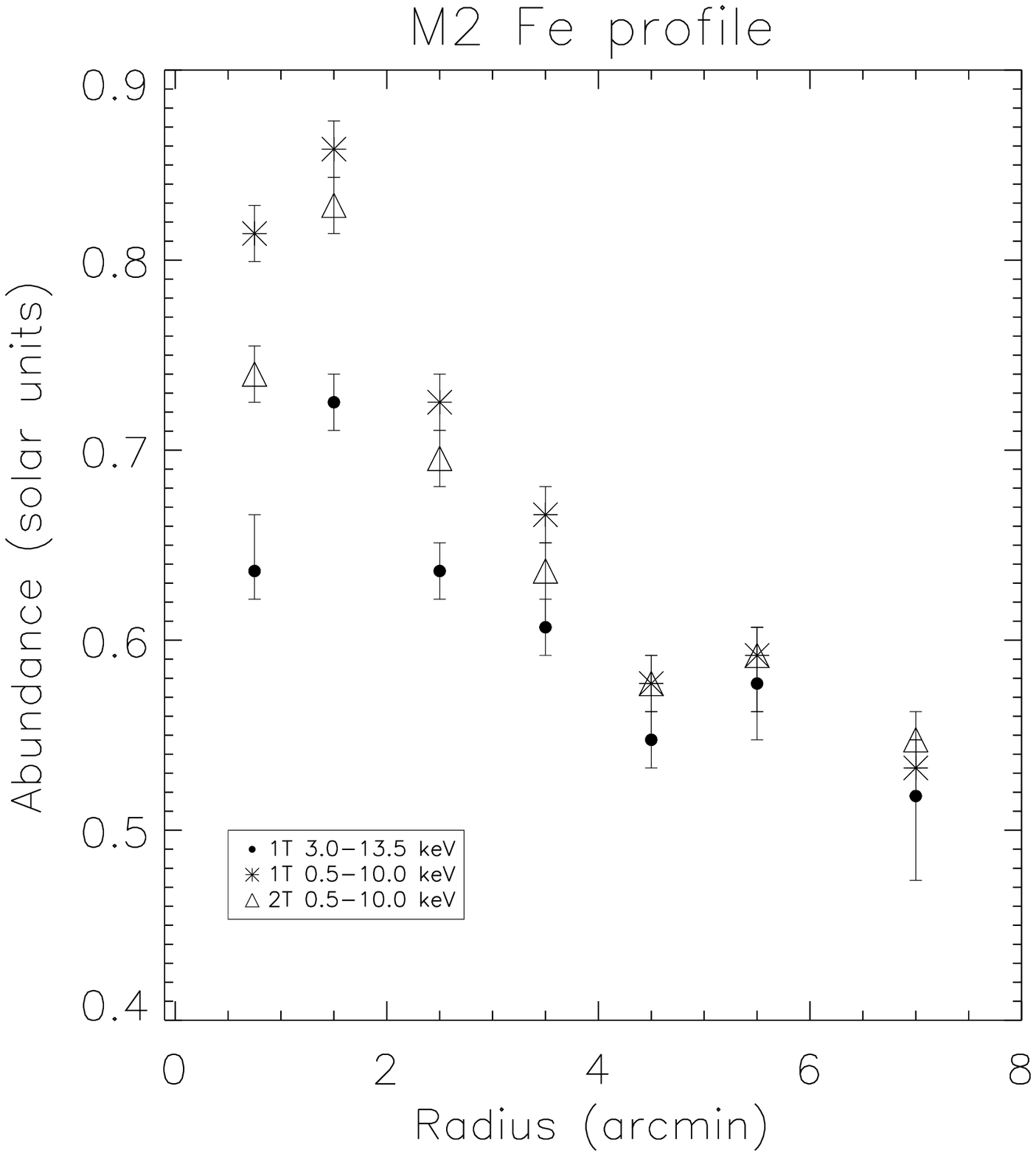}
\caption
{Fe abundance profiles obtained in the entire 0.5-10 keV band using one temperature (crosses) and two temperature models (open triangles) and, for comparison, the one obtained with the best fit model in the hard band (full circles), for PN singles data, on the top, and for MOS2 data, on the bottom.
Uncertainties are at the 68\% level for one interesting parameter
($\Delta \chi^{2}\,=\,1$). \label{fe2t}}
\end{figure}

\clearpage

\begin{figure}
\epsscale{1.3}
\plottwo{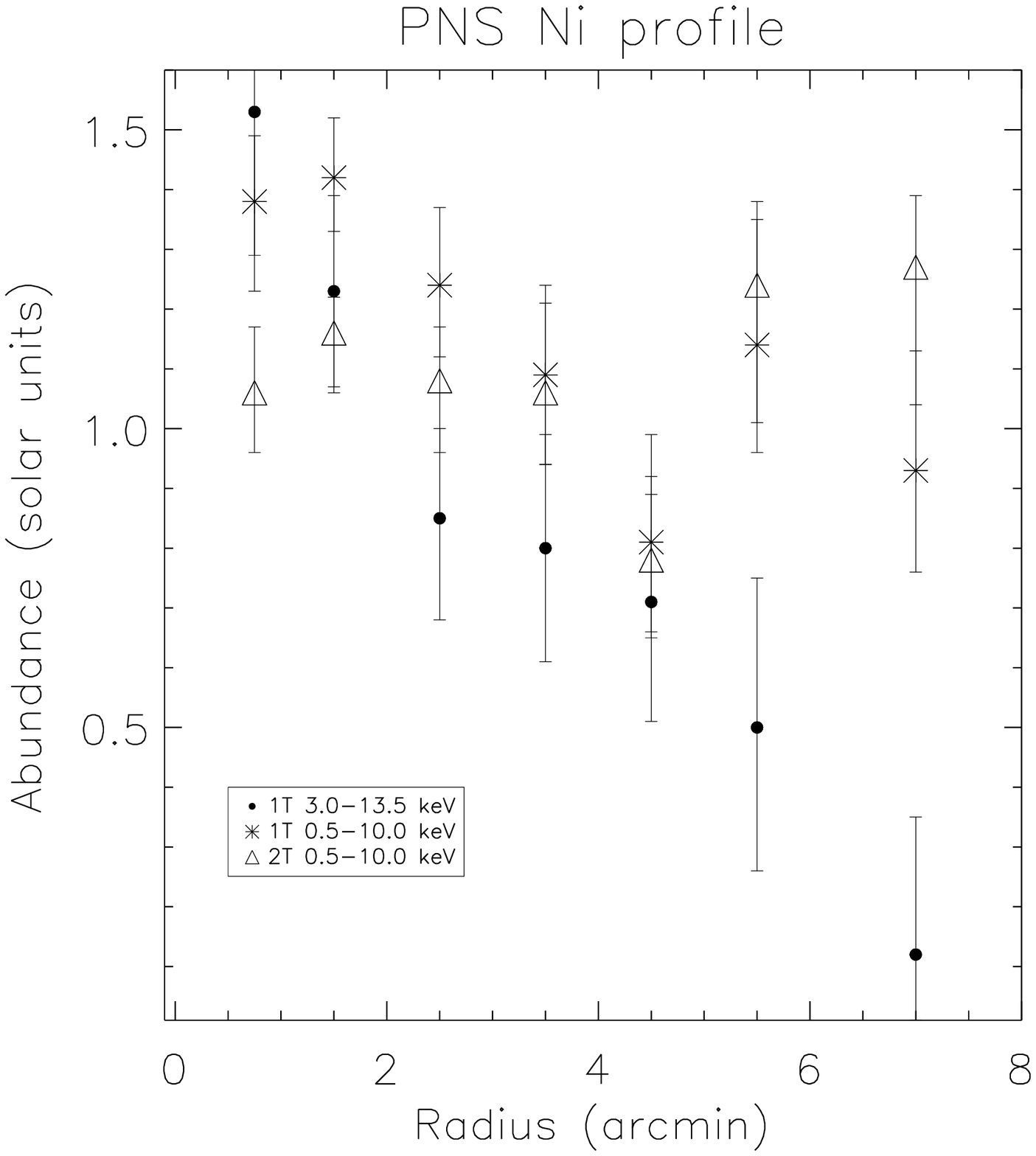}{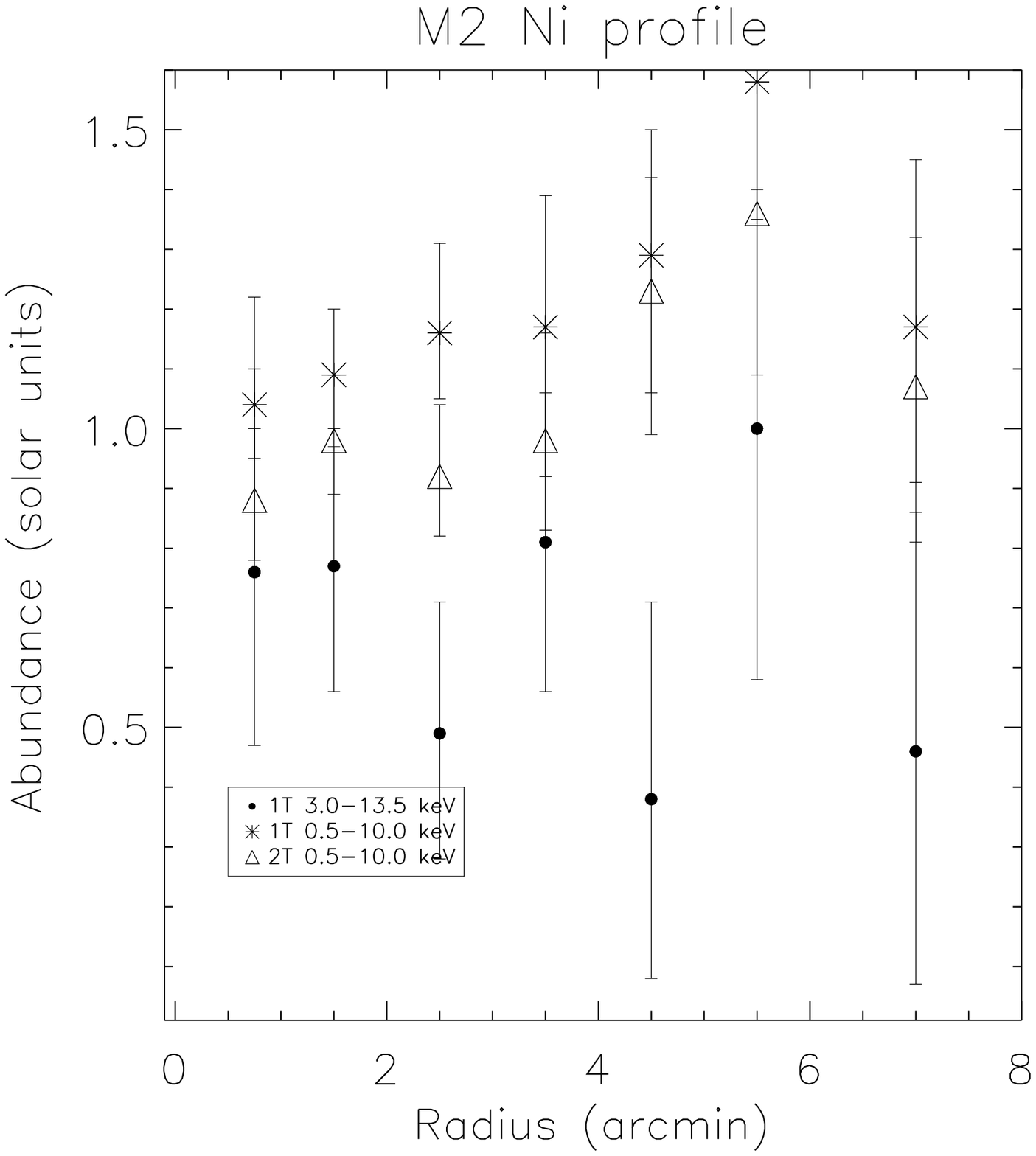}
\caption
{Ni abundance profiles obtained in the entire 0.5-10 keV band using one temperature (crosses) and two temperature models (open triangles) and, for comparison, the one obtained with the best fit model in the hard band (full circles), for PN singles data, on the top, and for MOS2 data, on the bottom. Uncertainties are at the 68\% level for one interesting parameter ($\Delta \chi^{2}\,=\,1$). \label{ni2t}}
\end{figure}

\clearpage

\begin{deluxetable}{cccccccccc}
\tablecolumns{10}
\tabletypesize{\footnotesize}
\tablewidth{0pt}
\rotate
\tablecaption{Parameters values for one and two temperatures models in different energy bands for the inner rings of the Perseus cluster using the 
single events of the PN camera. All errors quoted are at the 68\% level for one interesting parameter ($\Delta\chi^{2} = 1$
)
\label{pntab}}
\tablehead{
\colhead{Bin}  & \colhead{Mod-Band} & \colhead{$\rm{N_{H}}$} & \colhead{$\rm{kT_{h}}$} & \colhead{$\rm{EM_{h}}$} & \colhead{$\rm{kT_{c}}$} & \colhead{$\rm{EM_{c}}$} & \colhead{Fe} & \colhead{Ni} & \colhead{$\chi^{2}$/d.o.f}} 
\startdata
0.5$^{\prime}$-1$^{\prime}$ &  1T+pow/b (3-13.5 keV) &  & $3.91^{+0.11}_{-0.04}$ & $4.99^{+0.03}_{-0.09}$ & & &  $0.62^{+0.01}_{-0.01}$ & $1.53^{+0.27}_{-0.24}$ & 845/809 \\ 
                            & 1T (3-10 keV)         &  & $3.99^{+0.05}_{-0.06}$ & $4.91^{+0.02}_{-0.03}$ & & &  $0.62^{+0.01}_{-0.01}$ & $1.48^{+0.27}_{-0.24}$ & 834/799 \\
                            & 1T (0.5-10 keV)       & $1.09^{+0.01}_{-0.01}$ & $3.29^{+0.01}_{-0.01}$ & $5.75^{+0.01}_{-0.01}$ & & & $0.80^{+0.01}_{-0.01}$ & $1.38^{+0.11}_{-0.15}$ & 2037/1299 \\
                            &  2T (0.5-10 keV)       & $1.15^{+0.01}_{-0.01}$ & $5.00^{+0.08}_{-0.01}$ & $3.07^{+0.02}_{-0.03}$ & $2.03^{+0.02}_{-0.01}$ & $2.90^{+0.04}_{-0.07}$ & $0.69^{+0.01}_{-0.01}$ & $1.06^{+0.11}_{-0.10}$ & 1734/1297 \\

1$^{\prime}$-2$^{\prime}$ &  1T+pow/b (3-13.5 keV) &  & $4.01^{+0.04}_{-0.04}$ & $11.3^{+0.1}_{-0.1}$ & & &  $0.65^{+0.01}_{-0.01}$ & $1.23^{+0.16}_{-0.16}$ & 1114/1051 \\ 
                            & 1T (3-10 keV)         &  & $4.13^{+0.06}_{-0.02}$ & $11.2^{+0.1}_{-0.1}$ & & &  $0.64^{+0.01}_{-0.01}$ & $1.24^{+0.15}_{-0.17}$ & 1091/1020 \\
                            & 1T (0.5-10 keV)       & $1.12^{+0.01}_{-0.01}$ & $3.58^{+0.01}_{-0.01}$ & $12.19^{+0.01}_{-0.02}$ & & & $0.80^{+0.01}_{-0.01}$ & $1.42^{+0.10}_{-0.09}$ & 2894/1520 \\
                            &  2T (0.5-10 keV)       & $1.17^{+0.01}_{-0.01}$ & $4.95^{+0.02}_{-0.03}$ & $7.97^{+0.03}_{-0.05}$ & $2.04^{+0.01}_{-0.02}$ & $4.77^{+0.08}_{-0.03}$ & $0.71^{+0.01}_{-0.01}$ & $1.16^{+0.06}_{-0.10}$ & 2371/1518 \\

2$^{\prime}$-3$^{\prime}$ &  1T+pow/b (3-13.5 keV) &  & $4.86^{+0.07}_{-0.04}$ & $8.89^{+0.10}_{-0.13}$ & & &  $0.61^{+0.01}_{-0.01}$ & $0.85^{+0.15}_{-0.17}$ & 1079/1063 \\ 
                            & 1T (3-10 keV)         &  & $4.99^{+0.06}_{-0.04}$ & $8.83^{+0.07}_{-0.03}$ & & &  $0.61^{+0.01}_{-0.01}$ & $0.91^{+0.16}_{-0.17}$ & 1039/1020 \\
                            & 1T (0.5-10 keV)       & $1.14^{+0.01}_{-0.01}$ & $4.28^{+0.01}_{-0.02}$ & $9.74^{+0.01}_{-0.04}$ &  & & $0.72^{+0.01}_{-0.01}$ & $1.24^{+0.13}_{-0.12}$ & 2349/1519 \\
                            &  2T (0.5-10 keV)       & $1.18^{+0.01}_{-0.01}$ & $5.95^{+0.04}_{-0.03}$ & $7.22^{+0.03}_{-0.06}$ & $2.11^{+0.02}_{-0.01}$ & $2.86^{+0.08}_{-0.02}$ & $0.67^{+0.01}_{-0.01}$ & $1.08^{+0.09}_{-0.12}$ & 1978/1517 \\

3$^{\prime}$-4$^{\prime}$ &  1T+pow/b (3-13.5 keV) &  & $5.52^{+0.08}_{-0.07}$ & $6.44^{+0.04}_{-0.04}$ & & &  $0.56^{+0.01}_{-0.01}$ & $0.80^{+0.19}_{-0.19}$ & 1061/1080 \\
                          & 1T (3-10 keV)         &  & $6.03^{+0.08}_{-0.09}$ & $6.36^{+0.03}_{-0.03}$ & & &  $0.56^{+0.01}_{-0.01}$ & $0.87^{+0.16}_{-0.17}$ & 1062/1027 \\
                          & 1T+pow/b (0.5-10 keV)       & $1.15^{+0.01}_{-0.01}$ & $4.99^{+0.02}_{-0.03}$ & $6.93^{+0.01}_{-0.02}$ &  & & $0.61^{+0.01}_{-0.01}$ & $1.09^{+0.15}_{-0.15}$ & 2040/1527 \\

                          &  2T+pow/b (0.5-10 keV)       & $1.19^{+0.01}_{-0.01}$ & $5.64^{+0.02}_{-0.04}$ & $6.25^{+0.03}_{-0.02}$ & $1.64^{+0.03}_{-0.01}$ & $0.79^{+0.01}_{-0.04}$ & $0.56^{+0.03}_{-0.01}$ & $1.06^{+0.15}_{-0.12}$ & 1914/1525 \\ 

4$^{\prime}$-5$^{\prime}$ &  1T+pow/b (3-13.5 keV) &  & $5.97^{+0.08}_{-0.09}$ & $5.67^{+0.08}_{-0.06}$ & & &  $0.52^{+0.01}_{-0.01}$ & $0.71^{+0.21}_{-0.20}$ & 1210/1090 \\
                          & 1T+pow/b (0.5-10 keV)       & $1.13^{+0.02}_{-0.01}$ & $5.57^{+0.02}_{-0.03}$ & $5.99^{+0.01}_{-0.02}$ &  & & $0.55^{+0.01}_{-0.01}$ & $0.81^{+0.18}_{-0.16}$ & 1960/1521 \\

                          &  2T+pow/b (0.5-10 keV)       & $1.19^{+0.01}_{-0.01}$ & $6.26^{+0.04}_{-0.03}$ & $5.49^{+0.03}_{-0.01}$ & $1.61^{+0.03}_{-0.04}$ & $0.60^{+0.01}_{-0.03}$ & $0.53^{+0.03}_{-0.01}$ & $0.78^{+0.11}_{-0.12}$ & 1877/1519 \\ 

5$^{\prime}$-6$^{\prime}$ &  1T+pow/b (3-13.5 keV) &  & $6.36^{+0.14}_{-0.11}$ & $4.69^{+0.02}_{-0.03}$ & & &  $0.49^{+0.01}_{-0.01}$ & $0.50^{+0.25}_{-0.24}$ & 1095/1072 \\
                          & 1T+pow/b (0.5-10 keV)       & $1.13^{+0.01}_{-0.01}$ & $5.65^{+0.02}_{-0.05}$ & $5.06^{+0.01}_{-0.02}$ &  & & $0.50^{+0.01}_{-0.01}$ & $1.14^{+0.21}_{-0.18}$ & 2004/1497 \\

                          &  2T+pow/b (0.5-10 keV)       & $1.18^{+0.01}_{-0.01}$ & $6.36^{+0.06}_{-0.06}$ & $4.69^{+0.02}_{-0.01}$ & $1.37^{+0.03}_{-0.01}$ & $0.42^{+0.01}_{-0.03}$ & $0.47^{+0.02}_{-0.01}$ & $1.24^{+0.14}_{-0.23}$ & 1898/1495 \\

6$^{\prime}$-8$^{\prime}$ &  1T+pow/b (3-13.5 keV) &  & $6.78^{+0.13}_{-0.11}$ & $7.44^{+0.11}_{-0.20}$ & & &  $0.47^{+0.01}_{-0.01}$  & $0.12^{+0.23}_{-0.11}$ & 1249/1289 \\
                          & 1T+pow/b (0.5-10 keV)       & $1.15^{+0.01}_{-0.01}$ & $5.85^{+0.03}_{-0.04}$ & $8.04^{+0.01}_{-0.02}$ &  & & $0.49^{+0.03}_{-0.01}$ & $0.93^{+0.20}_{-0.17}$ & 2274/1637 \\

                          &  2T+pow/b (0.5-10 keV)       & $1.21^{+0.01}_{-0.01}$ & $6.69^{+0.04}_{-0.06}$ & $7.50^{+0.02}_{-0.02}$ & $1.29^{+0.01}_{-0.03}$ & $0.63^{+0.02}_{-0.02}$ & $0.47^{+0.01}_{-0.01}$ & $1.27^{+0.12}_{-0.23}$ & 2009/1635 \\
\enddata
\end{deluxetable}

\clearpage

\begin{deluxetable}{cccccccccc}
\tablecolumns{10}
\tabletypesize{\footnotesize}
\tablewidth{0pt}
\rotate
\tablecaption{Parameters values for one and two temperatures models in different energy bands for the inner rings of the Perseus cluster using the 
the MOS2 camera. All errors quoted are at the 68\% level for one interesting parameter ($\Delta\chi^{2} = 1$
)
\label{m2tab}}
\tablehead{
\colhead{Bin}  & \colhead{Mod-Band} & \colhead{$\rm{N_{H}}$} & \colhead{$\rm{kT_{h}}$} & \colhead{$\rm{EM_{h}}$} & \colhead{$\rm{kT_{c}}$} & \colhead{$\rm{EM_{c}}$} & \colhead{Fe} & \colhead{Ni} & \colhead{$\chi^{2}$/d.o.f}} 
\startdata
0.5$^{\prime}$-1$^{\prime}$ & 1T+pow/b (3-12 keV) &  & $3.84^{+0.06}_{-0.04}$ & $5.36^{+0.02}_{-0.03}$ & & &  $0.64^{+0.03}_{-0.01}$ & $0.76^{+0.34}_{-0.29}$ & 385/323 \\ 
                            & 1T (3-10 keV)         &  & $3.86^{+0.07}_{-0.03}$ & $5.37^{+0.02}_{-0.03}$ & & &  $0.64^{+0.03}_{-0.01}$ & $0.83^{+0.30}_{-0.30}$ & 372/318 \\
                            & 1T (0.5-10 keV)       & $1.17^{+0.01}_{-0.01}$ & $3.38^{+0.01}_{-0.02}$ & $5.91^{+0.01}_{-0.01}$ & & & $0.81^{+0.01}_{-0.01}$ & $1.04^{+0.18}_{-0.09}$ & 810/479 \\
                            &  2T (0.5-10 keV)       & $1.20^{+0.02}_{-0.01}$ & $5.18^{+0.05}_{-0.06}$ & $2.82^{+0.03}_{-0.09}$ & $2.33^{+0.02}_{-0.02}$ & $3.26^{+0.25}_{-0.08}$ & $0.74^{+0.01}_{-0.01}$ & $0.88^{+0.12}_{-0.10}$ & 669/477 \\

1$^{\prime}$-2$^{\prime}$ &  1T+pow/b (3-12 keV) &  & $4.14^{+0.05}_{-0.06}$ & $12.2^{+0.1}_{-0.1}$ & & &  $0.72^{+0.01}_{-0.01}$ & $0.77^{+0.20}_{-0.21}$ & 436/393 \\ 
                            & 1T (3-10 keV)         &  & $4.20^{+0.05}_{-0.05}$ & $12.1^{+0.1}_{-0.2}$ & & &  $0.71^{+0.01}_{-0.01}$ & $0.80^{+0.21}_{-0.20}$ & 435/379 \\
                            & 1T (0.5-10 keV)       & $1.18^{+0.01}_{-0.01}$ & $3.67^{+0.01}_{-0.01}$ & $13.00^{+0.01}_{-0.02}$ & & & $0.86^{+0.01}_{-0.01}$ & $1.09^{+0.11}_{-0.09}$ & 1228/540 \\
                            &  2T (0.5-10 keV)       & $1.19^{+0.01}_{-0.01}$ & $6.68^{+0.06}_{-0.09}$ & $4.09^{+0.03}_{-0.14}$ & $2.91^{+0.01}_{-0.01}$ & $9.15^{+0.06}_{-0.05}$ & $0.83^{+0.01}_{-0.01}$ & $0.98^{+0.11}_{-0.09}$ & 966/538 \\

2$^{\prime}$-3$^{\prime}$ &  1T+pow/b (3-12 keV) &  & $4.96^{+0.06}_{-0.11}$ & $10.00^{+0.15}_{-0.05}$ & & &  $0.64^{+0.01}_{-0.01}$ & $0.49^{+0.22}_{-0.21}$ & 406/402 \\
                          & 1T (3-10 keV)         &  & $5.17^{+0.05}_{-0.09}$ & $9.83^{+0.12}_{-0.07}$ & & &  $0.65^{+0.01}_{-0.02}$ & $0.59^{+0.24}_{-0.20}$ & 415/383 \\
                          & 1T (0.5-10 keV)       & $1.17^{+0.01}_{-0.01}$ & $4.35^{+0.02}_{-0.01}$ & $10.96^{+0.02}_{-0.01}$ &  & & $0.72^{+0.01}_{-0.01}$ & $1.16^{+0.15}_{-0.11}$ & 1106/544 \\

                          &  2T (0.5-10 keV)       & $1.21^{+0.01}_{-0.01}$ & $5.78^{+0.05}_{-0.03}$ & $7.84^{+0.05}_{-0.07}$ & $2.32^{+0.03}_{-0.02}$ & $3.43^{+0.08}_{-0.06}$ & $0.69^{+0.01}_{-0.01}$ & $0.92^{+0.12}_{-0.10}$ & 841/542 \\

3$^{\prime}$-4$^{\prime}$ &  1T+pow/b (3-12 keV) &  & $5.78^{+0.09}_{-0.10}$ & $7.54^{+0.09}_{-0.07}$ & & &  $0.61^{+0.01}_{-0.01}$ & $0.81^{+0.26}_{-0.25}$ & 407/406 \\
                          & 1T (3-10 keV)         &  & $6.15^{+0.09}_{-0.11}$ & $7.37^{+0.06}_{-0.05}$ & & &  $0.62^{+0.01}_{-0.02}$ & $1.03^{+0.26}_{-0.25}$ & 415/380 \\
                          & 1T+pow/b (0.5-10 keV)       & $1.17^{+0.02}_{-0.01}$ & $5.25^{+0.04}_{-0.02}$ & $7.93^{+0.01}_{-0.01}$ &  & & $0.67^{+0.01}_{-0.01}$ & $1.17^{+0.22}_{-0.15}$ & 782/542 \\

                          &  2T+pow/b (0.5-10 keV)       & $1.21^{+0.01}_{-0.01}$ & $6.30^{+0.05}_{-0.10}$ & $5.31^{+0.03}_{-0.08}$ & $2.74^{+0.05}_{-0.11}$ & $1.47^{+0.14}_{-0.06}$ & $0.64^{+0.01}_{-0.02}$ & $0.98^{+0.18}_{-0.15}$ & 687/540 \\

4$^{\prime}$-5$^{\prime}$ &  1T+pow/b (3-12 keV) &  & $6.14^{+0.12}_{-0.17}$ & $6.09^{+0.06}_{-0.07}$ & & &  $0.55^{+0.01}_{-0.01}$ & $0.38^{+0.33}_{-0.30}$ & 412/401 \\
                          & 1T+pow/b (0.5-10 keV)       & $1.26^{+0.01}_{-0.02}$ & $5.60^{+0.03}_{-0.04}$ & $6.37^{+0.01}_{-0.02}$ &  & & $0.58^{+0.01}_{-0.01}$ & $1.29^{+0.21}_{-0.23}$ & 701/538 \\

                          &  2T+pow/b (0.5-10 keV)       & $1.27^{+0.02}_{-0.01}$ & $6.66^{+0.04}_{-0.11}$ & $6.62^{+0.05}_{-0.07}$ & $2.36^{+0.06}_{-0.07}$ & $1.13^{+0.12}_{-0.04}$ & $0.58^{+0.01}_{-0.02}$ & $1.23^{+0.19}_{-0.24}$ & 666/536 \\

5$^{\prime}$-6$^{\prime}$ &  1T+pow/b (3-12 keV) &  & $6.67^{+0.23}_{-0.20}$ & $4.83^{+0.06}_{-0.05}$ & & &  $0.58^{+0.03}_{-0.03}$ & $1.00^{+0.40}_{-0.42}$ & 363/382 \\
                          & 1T+pow/b (0.5-10 keV)       & $1.16^{+0.01}_{-0.01}$ & $6.17^{+0.05}_{-0.07}$ & $5.01^{+0.02}_{-0.01}$ &  & & $0.59^{+0.01}_{-0.03}$ & $1.58^{+0.28}_{-0.33}$ & 650/518 \\

                          &  2T+pow/b (0.5-10 keV)       & $1.18^{+0.02}_{-0.02}$ & $7.11^{+0.07}_{-0.19}$ & $4.22^{+0.05}_{-0.06}$ & $3.31^{+0.15}_{-0.18}$ & $0.89^{+0.12}_{-0.09}$ & $0.59^{+0.01}_{-0.03}$ & $1.36^{+0.32}_{-0.27}$ & 639/516 \\

6$^{\prime}$-8$^{\prime}$ &  1T+pow/b (3-12 keV) &  & $7.20^{+0.21}_{-0.21}$ & $7.88^{+0.09}_{-0.09}$ & & & $0.52^{+0.03}_{-0.03}$  & $0.46^{+0.40}_{-0.39}$ & 447/475 \\
                          & 1T+pow/b (0.5-10 keV)       & $1.20^{+0.02}_{-0.01}$ & $6.50^{+0.06}_{-0.04}$ & $8.27^{+0.01}_{-0.02}$ &  & & $0.53^{+0.01}_{-0.01}$ & $1.17^{+0.28}_{-0.26}$ & 848/572 \\

                          &  2T+pow/b (0.5-10 keV)       & $1.23^{+0.02}_{-0.02}$ & $7.98^{+0.09}_{-0.14}$ & $6.95^{+0.06}_{-0.18}$ & $2.74^{+0.07}_{-0.14}$ & $1.46^{+0.23}_{-0.03}$ & $0.55^{+0.01}_{-0.01}$ & $1.07^{+0.25}_{-0.26}$ & 798/570 \\

\enddata
\end{deluxetable}


\begin{thebibliography}{}

\bibitem[Anders \& Grevesse(1989)]{anders89} Anders, E. \& Grevesse, N. 1989, Geochimica et Cosmochimica Acta, 53, 197

\bibitem[Arnaud(1996)]{arnaud96} Arnaud, K.A., 1996, Astronomical Data Analysis Software and Systems V, eds. Jacoby G. and Barnes J., p17, ASP Conf. Series volume 101 

\bibitem[Arnaud et al.(2001)]{arnaud01} Arnaud, M., Neumann, D. M., Aghanim, N., Gastaud, R., Majerowicz, S., Hughes, J. P. 2001, A\&A, 365, L80

\bibitem[Caon et al.(2000)]{caon00} Caon, N., Macchetto, d., Pastoriza, M., 2000, ApJS, 127, 39

\bibitem[Churazov(2003)]{churazov03} Churazov, E., 2003, presentation held at the conference The Riddle of Cooling Flows in Galaxies and Clusters of Galaxies, Charlottesville, VA, USA. May 31 -- June 4, 2003, ${\tt http://www.astro.virginia.edu/coolflow/abs.php}$

\bibitem[Costa et al.(2001)]{costa01} Costa, E., Soffitta, P., Bellazzini, R., Brez, A., Lumb, N., Spandre, G., 2001, Nature, 411, 662

\bibitem[De Grandi \& Molendi(2002a)]{grandi02a} De Grandi, S. \& Molendi, S., 2002a, ApJ, 567, 163

\bibitem[De Grandi \& Molendi(2002b)]{grandi02b} De Grandi, S. \& Molendi, S., 2002b, in Chemical Enrichment of Intracluster and Intergalactic Medium, ASP Conference Proceedings Edited by Roberto Fusco-Femiano and Francesca Matteucci. Vol 253, p.3

\bibitem[Dupke \& Arnaud(2001)]{dupke01} Dupke, R. A. \& Arnaud, K.A., 2001, ApJ, 548, 141

\bibitem[Fabian et al.(2002)]{fabian02} Fabian, A.C., Celotti, A., Blundell, K.M., Kassim, N.E., Perley, R.A., 2002, MNRAS, 331, 369

\bibitem[Fabian et al.(2003a)]{fabian03a} Fabian, A.C., Sanders, J.S., Allen, S.W., Crawford, C.S., Iwasawa, K., Johnstone, R.M., Schmidt, R.W., Taylor, G.B., 2003a, MNRAS in press, (astro-ph/030636)

\bibitem[Fabian et al.(2003b)]{fabian03b} Fabian, A.C., Sanders, J.S., Crawford, C.S., Conselice, C.J., Gallagher III, J.S., Wyse, R.F.G., 2003b, MNRAS in press, (astro-ph/030639)

\bibitem[Finoguenov et al.(2002)]{fino02} Finoguenov, A., Matsushita, K., B\"ohringer, H., Ikebe, Y., Arnaud, M., 2002, A\&A, 381, 21

\bibitem[Freyberg et al.(2002)]{freyberg02} Freyberg, M.J., Briel, U.G., Dennerl, K., Haberl, F., Hartner, G., Kendziorra, E., Kirsch, M., 2002, in Symp. New visions of the X-ray Universe in the XMM-Newton and Chandra era (ESA SP-488; Noordwijk: ESA)

\bibitem[Fukazawa et al.(2000)]{fukazawa00} Fukazawa, Y., Makishima, K., Tamura, T., Nakazawa, K., Ezawa, H., Ikebe, Y., Kikuchi, K., Ohashi, T., 2000, MNRAS, 313, 21

\bibitem[Gastaldello \& Molendi(2002)]{gasta02} Gastaldello, F. \& Molendi, S., 2002, ApJ, 572, 160

\bibitem[Gastaldello et al.(2003)]{gasta03} Gastaldello, F., Ettori, S., Molendi, S., Bardelli, S., Venturi, T., Zucca, E., 2003, A\&A in press, (astro-ph/0307342)

\bibitem[Gilfanov et al.(1987)]{gilfanov87} Gilfanov, M.R., Sunyaev,R.A., Churazov, E.M., 1987, Sov. Astron. Lett., 13, 3

\bibitem[Grupe(2001)]{grupe01} Grupe, D., 2001, ${\tt http://wave.xray.mpe.mpg.de/xmm/cookbook/EPIC\_PN/ootevents.html}$ 

\bibitem[Grevesse \& Sauval(1998)]{grevesse98} Grevesse, N. \& Sauval, A. J., 1998, Space Science Reviews, 85, 161

\bibitem[Kaastra(1992)]{kaastra92} Kaastra, J.S., 1992, An X-ray Spectral code for Optically Thin Plasmas (Internal SRON-Leiden Report, updated version 2.0)

\bibitem[Kaiser(2003)]{kaiser03} Kaiser, C.R., 2003, MNRAS in press, (astro-ph/0305104)

\bibitem[Kirsch et al.(2002)]{kirsch02} Kirsch, M. et al., ``Status of the EPIC calibration and data analysis'', 2002, XMM-SOC-CAL-TN-018

\bibitem[Jansen et al.(2001)]{jansen01} Jansen, F., Lumb, D., Altieri, B., Clavel, J., Ehle, M., Erd, C., Gabriel, C., Guainazzi, M., Gondoin, P., Much, R., Munoz, R., Santos, M., Schartel, N., Texier, D., Vacanti, G. 2001, A\&A, 365, L1

\bibitem[Liedahl et al.(1995)]{liedahl95} Liedahl, D. A., Osterheld A. L., Goldstein, W. H. 1995, ApJ, 438, L115 

\bibitem[Lumb(2002)]{lumb02} Lumb, D., ``EPIC background files'', 2002, XMM-SOC-CAL-TN-016

\bibitem[Marty et al.(2002)]{marty02} Marty, P.B., Kneib, J.P., Sadat, R., Ebeling, H., Smail, I., 2002, proceedings SPIE vol. 4851 (astro-ph/0209270) 

\bibitem[Mathews et al.(2001)]{mathews01} Mathews, W.G., Buote, D.A., Brighenti, F., 20001, ApJ, 550, L31

\bibitem[Matsushita et al.(2002)]{matsu02} Matsushita, K., Belsole, E., Finoguenov, A., B\"ohringer, H., 2002, A\&A, 386, 77

\bibitem[Mewe et al.(1985)]{mewe85} Mewe, R., Gronenschild, E. H. B. M., van den Oord, G. H. J., 1985, A\&AS, 62, 197

\bibitem[Molendi et al.(1998)]{mole98} Molendi, S., Matt, G., Antonelli, L.A., Fiore, F., Fusco-Femiano, R., Kaastra, J., Maccarone, C., Perola, C., 1998, ApJ, 499, 618

\bibitem[Molendi et al.(1999)]{mole99} Molendi, S., De Grandi, S., Fusco-Femiano, R., Colafrancesco, S., Fiore, F., Nesci, R., Tamburelli, F., 1999, ApJ, 525, L73 

\bibitem[Molendi(2002)]{mole02} Molendi, S., 2002, ApJ, 580, 815

\bibitem[Molendi \& Sembay(2003)]{mole03} Molendi, S. \& Sembay, S., 2003, XMM-SOC-CAL-TN-0036

\bibitem[Nevalainen et al.(2003)]{neva03} Nevalainen, J., Lieu, R., Bonamente, M., Lumb, D., 2003, ApJ, 584, 716 

\bibitem[Nomoto et al.(1997)]{nomoto97} Nomoto, K., Iwamoto, K., Nakasato, N., Thielemann, F. K., Brachwitz, F., Tsujimoto, T., Kubo, Y., Kishimoto, N. 1997, Nucl.Phys. A, 621, 467

\bibitem[Peres et al.(1998)]{peres98} Peres, C.B., Fabian, A.C., Edge, A.C., Allen, S.W., Johnstone, R.M., White, D.A., 1998, MNRAS, 298, 416

\bibitem[Pratt \& Arnaud(2002)]{pratt02} Pratt, G.W. \& Arnaud, M., 2002, A\&A, 394, 375

\bibitem[Sakelliou et al.(2002)]{sakelliou02} Sakelliou, I., Peterson, J.R., Tamura, T., Paerels, F.B.S., Kaastra, J.S., Belsole, E., B\"ohringer, H., Branduardi-Raymont, G., Ferrigno, C., den Herder, J.W., Kennea, J., Mushotzky, R.F., Vestrand, W.T., Worrall, D.M., 2002, A\&A, 391, 903

\bibitem[Saxton \& Siddiqui(2002)]{saxton02} Saxton, R.D. \& Siddiqui, H., ``The status
of the SAS spectral response generation tasks for XMM-EPIC'', 2002, XMM-SOC-PS-TN-43 

\bibitem[Sazonov et al.(2002)]{sazonov02} Sazonov, S.Yu., Churazov, E.M., Sunyaev, R.A., 2002, MNRAS, 333, 191

\bibitem[Schmidt et al.(2002)]{schmidt02} Schmidt, R.W., Fabian, A.C., Sanders, J.S., 2002, MNRAS, 337, 71 

\bibitem[Smith et al.(2001)]{smith01} Smith, R. K., Brickhouse, N. S., Liedahl, D. A., Raymond, J. C. 2001, ApJ, 556, L91

\bibitem[Sparks et al.(1989)]{sparks89} Sparks, W.B., Macchetto, D., Golombeck, D., 1989, ApJ, 345, 153

\bibitem[Tamura et al.(2003)]{tamura03} Tamura, T., Kaastra, J.S., Makishima, K., Takahashi, I., 2003, A\&A, 399, 407 

\bibitem[Xu et al.(2002)]{xu02} Xu, H., Kahn, S.M., Peterson, J.R., Behar, E., Paerels, F.B.S., Mushotzky, R.F., Jernigan, J.G., Brinkman, A.C., Makishima, K., 2002, ApJ, 579, 600

\end{thebibliography}
\end{document}